\documentclass[11pt]{article}
\usepackage[margin=1in]{geometry}
\usepackage[utf8]{inputenc}
\usepackage{graphicx}
\usepackage{qcircuit}
\usepackage{tikz}
\usetikzlibrary{matrix, arrows.meta}
\usepackage{subcaption}
\usepackage{float}
\usepackage{amsmath, amssymb, amsthm}
\usepackage[ruled, vlined, linesnumbered]{algorithm2e}
\usepackage{cite}
\usepackage{multirow}
\usepackage{hyperref}
\usepackage{authblk}
\usepackage{xcolor}
\usepackage[normalem]{ulem}

\newcommand{\bs}[1]{\boldsymbol #1}
\newtheorem{definition}{Definition}

\title{\bf Phase polynomials synthesis algorithms for NISQ architectures and beyond}
\date{}
\author[1]{Vivien Vandaele}
\author[1]{Simon Martiel}
\author[2]{Timoth\'ee Goubault de Brugi\`ere}
\affil[1]{Atos Quantum Lab, Les Clayes-sous-bois, France}
\affil[2]{Laboratoire de Recherche en Informatique, Orsay, France}

\begin{document}
\maketitle

\begin{abstract}
We present a framework for the synthesis of phase polynomials that addresses both cases of full connectivity and partial connectivity for NISQ architectures.
In most cases, our algorithms generate circuits with lower CNOT count and CNOT depth than the state of the art or have a significantly smaller running time for similar performances.
We also provide methods that can be applied to our algorithms in order to trade an increase in the CNOT count for a decrease in execution time, thereby filling the gap between our algorithms and faster ones.
\end{abstract}

\section{Introduction}
Quantum circuits optimization is essential to foster the practicability and efficiency of quantum computation.
In particular, to cope with the much-needed compactness of quantum circuits, the synthesis of reversible circuits is being studied thoroughly.
Because the $T$ gate has a high fault-tolerant implementation cost~\cite{vuillot}, much work has been put into the minimization of the $T$-count~\cite{kissinger, amy_2019, heyfron, gosset, beaudrap, mosca_count, beaudrap_2020, zhang2019optimizing} and the $T$-depth~\cite{amy_2013, tpar, abdessaied2014quantum, mosca_depth}.
In contrast, the CNOT gate has a low implementation cost as it is part of the Clifford group~\cite{gottesman}.
Nonetheless, the usage of metrics based on the $T$ gate have limitations, it turns out that the number of CNOT gates in a circuit is a metric that should not be overlooked as it can have a significant impact on the implementation cost of a circuit~\cite{maslov}.

On top of that, quantum computers in the Noisy Intermediate Scale Quantum (NISQ) era~\cite{preskill} have architectural constraints.
Concretely, the qubits within these computers are not connected in an all-to-all manner.
It implies that logical gates having an arity of $2$, such as CNOT gates, can only be applied between certain pairs of qubits.
Thus, making a circuit compliant with a given architecture inevitably causes an increase in the CNOT count~\cite{linke}.

A common way of dealing with architectural constraints is to insert SWAP gates to route logical qubits~\cite{wille, hirata, sabre, childs}.
An alternative is to perform architecture-aware synthesis~\cite{griend}, a method which often produces circuits with a much lower CNOT count while satisfying the architectural constraints.
This approach is typically applied on subsets of circuits that can be represented by high-level constructs such as linear reversible functions.
These circuits can then be put together to form a complete architecture compliant quantum circuit~\cite{lazy, mosca}.
An important building block in this compilation scheme is the synthesis of circuits composed exclusively of CNOT and $R_Z$ gates.
These circuits can be represented by a high-level construct called phase polynomials.
In this work we tackle the phase polynomials synthesis problem and propose efficient algorithms for both cases of restricted and complete connectivity.\\

\noindent{\bf State of the art.}
In~\cite{sat}, a SAT-based algorithm that optimally solves the phase polynomials synthesis problem is proposed.
Although this method offers good results regarding the CNOT count, it has an exponential complexity since the SAT problem is NP-complete and is therefore only practical for the synthesis of small phase polynomials.
An efficient heuristic algorithm for phase polynomials synthesis is provided by Amy et al. in~\cite{gray}.
This algorithm, named Gray-Synth, is inspired by Gray code~\cite{frank1953pulse} and is considered as the current state of the art.
There exists numerous other algorithms for phase polynomials synthesis.
Some of them don't have CNOT minimization as primary objective, as it is the case of the $T$par algorithm~\cite{tpar} that aims to parallelize the phase gates of a phase polynomial.
This is also the case, with a lower degree, in~\cite{nam} where automated methods for the optimization of large quantum circuit are given. 
As their algorithm scales similarly to the Gray-Synth algorithm and is purposely designed for specific circuits, we will prefer to compare our algorithm with the Gray-Synth algorithm.

Regarding the complexity of the problem, the results presented in~\cite{herr} lead us to think that it is intractable.
The same authors of the Gray-Synth algorithm corroborate this idea by proving the NP-completeness of the problem in some restricted cases~\cite{gray}.

Qubit routing could be used to make the Gray-Synth algorithm compliant with constrained architectures.
This idea was developed and greatly refined by Nash et al.~\cite{nash}, and a modified version of their algorithm has been implemented in the Staq toolkit~\cite{staq}.
An altered version of this algorithm was also recently incorporated by Gheorghiu et al. in a slice-and-build algorithm that optimizes a given quantum circuit while taking into account the connectivity constraints imposed by the physical hardware architecture~\cite{mosca}.
In a recent work and with a similar goal, a framework composed of greedy architecture-aware synthesis routines for the compilation of quantum circuits was presented in~\cite{lazy}.

A phase polynomial is partially composed of a set of parities which can be stored in a parity table.
As explained in~\cite{gray}, a circuit in which each parity occurs at least once is called a parity network and can be easily modified in order to implement the corresponding phase polynomials.
In all parity network synthesis algorithms the parities are synthesized in an established order, we refer to this order as the parity ordering.
In the Gray-Synth algorithm~\cite{gray} this ordering is inspired by Gray code.
Most of the parity network synthesis algorithms for arbitrary connectivity follow this idea and are based on the Gray-Synth algorithm.
Yet, while the parity ordering defined by the Gray-Synth algorithm is efficient for all-to-all connectivity, it may be unfitted for arbitrary connectivity.
Indeed, most Gray-Synth based algorithms for arbitrary connectivity are not taking the architecture into account when establishing the parity ordering~\cite{nash, staq, mosca}.
In other words, the choice of the next parity to synthesize is solely based on the parity table, without taking into account the underlying graph of the architecture.
An algorithm proposed by Arianne Meijer-van de Griend and Ross Duncan~\cite{duncan} aims to solve this shortcoming by recursively considering only non-cutting vertices of the underlying graph.
However this algorithm is still based on the parity ordering of the Gray-Synth algorithm which is foremostly designed for all-to-all connectivity.\\

\noindent{\bf Our approach.}
In this paper we present an efficient alternative to the Gray code inspired parity ordering that helds better results in both cases of all-to-all connectivity and constrained architectures.
In our approach, the parity ordering is defined by a two steps iterative process.
The first step consists in choosing a parity and the second one corresponds to the synthesis of the chosen parity, we iterate until all parities have been synthesized.
Here the parity choice is not bound to a parity ordering uniquely defined upon the parity table as it is the case for Gray code inspired methods, but can also take into account an arbitrary connectivity.
In fact, we will see that this method can be easily adapted to constrained architectures by relying on the commonly used notion of Steiner tree.
This extension to constrained architectures induces an important time cost, nevertheless we will present some techniques to significantly reduce the running time of our algorithm while preserving an important CNOT count reduction when compared to the state of the art.\\

\noindent{\bf Outline.}
This paper is organized as follows.
Section~\ref{sec:phase_polynomials_synthesis} introduces the circuit-polynomial correspondence for quantum circuits over the $\{CNOT, R_z\}$ gate set.
In Section~\ref{sec:alltoallconnectivity} we present our heuristic algorithm for the synthesis of phase polynomials in the case of full connectivity.
In Section~\ref{sec:partialconnectivity} we extend our algorithm for partial connectivity and we give methods to lower the complexity of our approach.
Benchmarks are given at the end of Sections~\ref{sec:alltoallconnectivity} and~\ref{sec:partialconnectivity}.

\section{Phase polynomials synthesis}
\label{sec:phase_polynomials_synthesis}

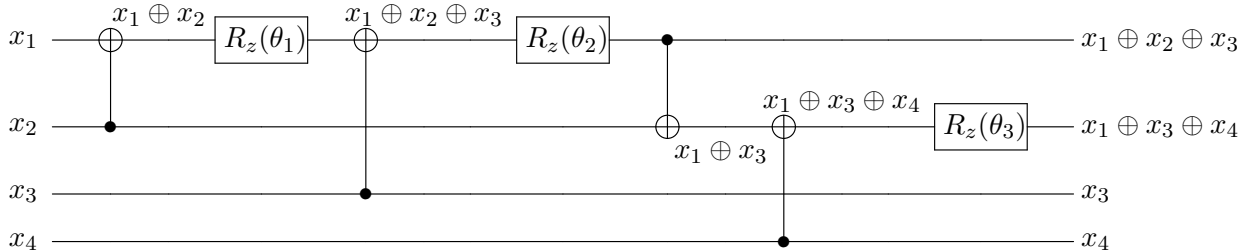
\begin{figure}[ht]
    {
    \centering
    \Qcircuit @C=1.6em @R=1.4em{
        & \lstick{x_1} &  \targ & \ustick{\!\!\!\!x_1\oplus x_2} \qw & \gate{\parbox{1.0cm}{\centering $R_z(\theta_1)$}} & \targ &\ustick{x_1\oplus x_2 \oplus x_3}\qw & \qw & \gate{\parbox{1.0cm}{\centering $R_z(\theta_2)$}} & \ctrl{1} & \qw & \qw & \qw & \qw & \qw & \qw & \rstick{\!\!\!\!\!\!\!\!\!\!\!x_1 \oplus x_2 \oplus x_3}\\
        & \lstick{x_2} & \ctrl{-1} & \qw & \qw & \qw & \qw & \qw & \qw & \targ & \dstick{\!\!x_1\oplus x_3}\qw & \targ & \ustick{x_1\oplus x_3 \oplus x_4}\qw & \qw & \gate{\parbox{1.0cm}{\centering $R_z(\theta_3)$}} & \qw & \rstick{\!\!\!\!\!\!\!\!\!\!\!x_1 \oplus x_3 \oplus x_4}\\
        & \lstick{x_3} & \qw & \qw & \qw & \ctrl{-2} & \qw & \qw & \qw & \qw & \qw & \qw & \qw & \qw & \qw & \qw & \rstick{\!\!\!\!\!\!\!\!\!\!\!x_3}\\
        & \lstick{x_4} & \qw & \qw & \qw & \qw & \qw & \qw & \qw & \qw & \qw & \ctrl{-2} & \qw & \qw & \qw & \qw & \rstick{\!\!\!\!\!\!\!\!\!\!\!x_4}\\
        & \\
    }
    }
\caption{An annotated circuit composed of CNOT and $R_z$ gates.}
\label{fig:annotated}
\end{figure}

Let $C$ be a quantum circuit operating over $n$ qubits and composed of CNOT and $R_z$ gates.
Such circuit can be best described by exploiting the circuit-polynomial correspondence~\cite{dawson, montanaro}, which associates a phase polynomial and a linear reversible function to $C$.
The action of $C$ on a basis state has the form 
$$\lvert \bs x\rangle \mapsto e^{2\pi ip(\bs x)} \lvert g(\bf \bs x)\rangle$$
where $g: \mathbb{F}_2^n \rightarrow \mathbb{F}_2^n$ is a linear reversible function and
$$ p(\bs x) = \sum_{i = 1}^{2^n} \theta_i\  f_{i}(\bs x)$$
is a linear combination of linear Boolean functions $f_{i}: \mathbb{F}_2^n \rightarrow \mathbb{F}_2$.
Any linear Boolean function $f_{i}$ can be written as 
$$f_{i}(\bs x) = \bs y^i \cdot \bs x= y^i_1x_1  \oplus y^i_2 x_2 \oplus \ldots y^i_n x_n$$
where $\bs y^i \in \mathbb{F}_2^n$ and $\oplus$ stands for the XOR operation.
The function $p(\bs x)$ is the phase polynomial associated with $C$, and we will refer to the Boolean vectors $\bs y^i$ as the parities of the phase polynomial $p(\bs x)$.
For instance, the circuit represented Figure~\ref{fig:annotated} performs the mapping 
$$ \lvert x_1, x_2, x_3, x_4 \rangle \mapsto e^{ip(x_1, x_2, x_3, x_4)} \lvert x_1 \oplus x_2 \oplus x_3, \; x_1 \oplus x_3 \oplus x_4,\; x_3,\; x_4 \rangle$$
where $p(x_1, x_2, x_3, x_4) = \theta_1(x_1 \oplus x_2) + \theta_2(x_1 \oplus x_2 \oplus x_3) + \theta_3(x_1 \oplus x_3 \oplus x_4)$.
The parities of a phase polynomial can be described by a matrix where each line represents a qubit and each column represents a parity having an associated angle not equal to $0$, we call this matrix the parity table of the phase polynomial and we denote it $P$.
In our example, the parity table of the phase polynomial is 
$$P =
    \label{matrix}
     \begin{pmatrix}
         1 & 1 & 1\\
         1 & 1 & 0\\
         0 & 1 & 1\\
         0 & 0 & 1\\
     \end{pmatrix}.
$$

Performing the synthesis of the phase polynomial $p(\bs x)$ and the linear reversible function $g(\bs x)$ amounts to constructing a circuit equivalent to $C$.
The synthesis of linear reversible functions is a well studied problem as there exists asymptotically optimal methods~\cite{pmh}, as well as efficient heuristic algorithms in both cases of partial and full connectivity~\cite{syndrome, tang}.
For that reason we will put aside the problem of synthesizing the linear reversible function $g(\bs x)$, and we will focus on the phase polynomials synthesis problem.\\

\noindent{\bf Parity networks.}
Following Amy et al.~\cite{gray}, we tackle the phase polynomials synthesis problem via the parity network formalism.
\begin{definition}[Parity network]
A parity network for a parity table $P$ is a CNOT circuit in which each parity $\bs y \in P$ appears at least once.
\end{definition}
The $R_z$ gate only modifies the phase and doesn't affect the logical value of the qubit.
Therefore, a parity network can be easily modified by exclusively inserting $R_z$ gates to implement any phase polynomial associated with the parity table $P$.
It implies that the phase polynomials synthesis problem can be reduced to the parity network synthesis problem.
For the remaining of this paper we will consider the parity network synthesis problem and ignore the rotation gates in order to focus on the core of the problem.

Recall that a CNOT gate performs the mapping $\lvert x_i \rangle\lvert x_j \rangle \mapsto \lvert x_i \rangle\lvert x_i \oplus x_j \rangle$.
When applying a CNOT$_{x_i, x_j}$ gate where $x_i$ is the control qubit and $x_j$ is the target qubit, the parity table can be expressed in the new basis by performing the row addition $P_i = P_i \oplus P_j$.
It follows that a parity $\bs y \in P$ is being carried out by a qubit if it satisfies $\sum_{i=1}^n y_i = 1$.

\section{Parity network synthesis for all-to-all connectivity}
\label{sec:alltoallconnectivity}
\subsection{An efficient heuristic algorithm}
\label{sec:alltoall}
In this section we formalize a heuristic algorithm for phase polynomials synthesis in all-to-all connectivity.
Our algorithm is presented in pseudo-code in Algorithm~\ref{code:alltoall} and an example is provided in Figure~\ref{ex:alltoall}.
The term CNOT$_{i,j}$ refers to a CNOT gate with control $i$ and target $j$ and we define the function $h$ as the Hamming weight of a binary vector or binary matrix, i.e. $h(\bs y) = \sum_{i=1}^n y_i$ where $\bs y \in \mathbb{F}_2^n$ and $h(P) = \sum_{\bs y \in P} h(\bs y)$ where $P$ is a parity table.
Let $S$ be a sequence of row additions, we denote by $\bs y^S$ (resp. $P^S$) the state of $\bs y$ (resp. $P$) after applying the sequence of additions in $S$ onto $\bs y$ (resp. $P$).
Our algorithm follows a two steps iterative process:
\begin{enumerate}
	\item Choose a parity $\bs y \in P$.
	\item Perform the synthesis of $\bs y$. Remove $\bs y$ from $P$ and go to step $1$.
\end{enumerate}
\noindent{\bf Step 1.}
To know which parity $\bs y \in P$ it would be judicious to choose in step $1$ of our algorithm we first reflect on the minimum CNOT cost induced by the synthesis of $\bs y$.
Let $S_{\bs y}$ be the set of minimum length sequences of additions such that for all $S \in S_{\bs y}$ we have $h(\bs y^S) = 1$.
In other words, $S_{\bs y}$ contains all minimum length sequences of additions that effectively synthesize $\bs y$.
Note that one addition can reduce the value of $h(\bs y)$ by at most $1$.
That being so, it is clear that the length of all $S \in S_{\bs y}$ is equal to $h(\bs y) - 1$, implying a minimum CNOT cost of $h(\bs y) - 1$ for the synthesis of $\bs y$.
Based on this fact it is rather intuitive to choose the parity $\bs y$ for which $h(\bs y)$ is minimum for the step $1$ of our algorithm.\\

\noindent{\bf Step 2.}
We are left with the second step of the algorithm that raises the following question: which sequence $S \in S_{\bs y}$ of additions should we choose to perform the synthesis of $\bs y$?
A natural choice would be to select the sequence $S \in S_{\bs y}$ that minimizes the value $h(P^S)$.
However we are faced with a challenge as the size of $S_{\bs y}$ is exponential with respect to $h(\bs y)$.
In fact, we will see that the size of $S_{\bs y}$ is greater than the number of spanning arborescences in a complete directed graph composed of $h(\bs y)$ vertices.
lWe define this graph as follows.

\begin{definition}[Parity graph]
	Let $P$ be a parity table and $\bs y$ be a parity of $P$, the parity graph associated with $\bs y$ is the complete directed graph $G_{\bs y}=(V, A)$ where $V=\{i \mid y_i = 1\}$ and where each arc $(i, j) \in A$ going from $i$ to $j$ is weighted by $w_{i,j} = h(P_i \oplus P_j) - h(P_j)$.
\end{definition}

Let $X$ be a spanning arborescence of $G_{\bs y}$, note that $X$ is composed of $h(\bs y) - 1$ arcs.
For each arc $(i, j) \in X$ going from $i$ to $j$ we associate the addition $P_j = P_j \oplus P_i$.
Now consider a successors-first traversal of $X$ as defined below.

\begin{definition}[Successors-first traversal]
	A traversal of an arborescence $X$ is a successors-first traversal if and only if for every vertex $i$ in $X$ the successors of $i$ in $X$ are visited before $i$.
\end{definition}

For example, the traversal resulting from a depth-first search postordering is a successors-first traversal.
We can construct a sequence $S$ of additions by following the order of this traversal: if $j$ is the currently visited vertex then we append to $S$ the addition associated with the unique arc $(i, j) \in X$ going from $i$ to $j$.
The length of $S$ is equal to $h(\bs y) - 1$ and $h(\bs y^S) = 1$, thus $S \in S_{\bs y}$ and we say that $X$ is the spanning arborescence associated with $S$.
A different successors-first traversal would give us a different sequence of additions for the same spanning arborescence $X$, and for every sequence $S \in S_{\bs y}$ there is a corresponding spanning arborescence in $G_{\bs y}$.
Hence there is a surjection between $S_{\bs y}$ and the set of spanning arborescences in $G_{\bs y}$, in particular we have $|S_{\bs y}| \geq |\{X \mid X\text{ is a spanning arborescence of }G_{\bs y}\}| = n^{n-1}$ where $n=h(\bs y)$.
As a result we have to choose $S\in S_{\bs y}$ among an exponential number of possibilities.

To cope with this problem we can first notice that if two sequences $S_i, S_j$ are associated with the same arborescence, then they are equivalent in the sense that $P^{S_i} = P^{S_j}$.
As our only metric is the number of CNOT we don't make any distinction between these equivalent sequences, and we can equivalently refer to $S \in S_{\bs y}$ or its associated spanning arborescence in $G_{\bs y}$.
Recall that each arc $(i, j) \in A$ going from $i$ to $j$ is weighted by $w_{i,j} = h(P_i \oplus P_j) - h(P_j)$, then we have $$h(P^S) = h(P) + \sum_{(i, j) \in X} h(P_i \oplus P_j) - h(P_j) = h(P) + \sum_{(i, j) \in X} w_{i,j} $$ where $S \in S_{\bs y}$ and $X$ is the spanning arborescence associated with $S$ in $G_{\bs y}$.
In consequence, minimizing $h(P^S)$ amounts to minimizing $\sum_{(i, j) \in X} w_{i, j}$, which is optimally satisfied when $X$ is the minimum weight spanning arborescence of $G_{\bs y}$.
The minimum weight spanning arborescence problem is a well-known problem, an algorithm proposed by Robert Endre Tarjan~\cite{tarjan} solves it with a complexity of $\mathcal{O}(|V|^2)$ for complete graphs where $|V|$ is the number of vertices.\\

\noindent{\bf Correctness and complexity.}
Let $P$ be a parity table of size $n \times m$.
Our algorithm terminates when $P$ is empty and a parity $\bs y \in P$ is removed from $P$ at each iteration.
Therefore our algorithm performs $m$ iterations and finishes.
The algorithm starts with an empty circuit $C$ and at each iteration $C$ is extended to synthesize a parity $\bs y \in P$ not yet occurring in $C$.
Hence the constructed circuit $C$ is a parity network for $P$ and our algorithm is correct.

Choosing the parity in the step $1$ of our algorithm has a cost of $\mathcal{O}(mn)$.
For the step $2$, constructing the graph $G_{\bs y}$ has a complexity of $\mathcal{O}(mn^2)$, whereas computing its minimum weight spanning arborescence and performing the for loop over the depth-first search both have a smaller complexity of $\mathcal{O}(n^2)$ and $\mathcal{O}(mn)$ respectively.
Both steps are performed $m$ times so the overall complexity of our algorithm is $\mathcal{O}(m^2n^2)$.\\

\begin{algorithm}[H]
	\SetAlgoLined
	\SetArgSty{textnormal}
	\SetKwInput{KwInput}{Input}
	\SetKwInput{KwOutput}{Output}
	\KwInput{Parity table $P$}
	\KwOutput{Circuit synthesizing a parity network associated with $P$}
	$C \leftarrow$ new empty circuit\\
	\While{$P$ \text{non-empty}}{
		$\displaystyle \bs y \leftarrow \min \operatorname*{argmin}_{\bs y} \{h(\bs y) \mid \bs y \in P\}$\\
		$P \leftarrow P \setminus \bs y$\\
		$G_{\bs y} \leftarrow (\{i \mid y_i = 1\}, \{(i, j, h(P_i \oplus P_j) - h(P_j)) \mid y_i=1, y_j=1, i \neq j\})$\\
		$X \leftarrow$ MinimumWeightSpanningArborescence$(G_{\bs y})$\\
		\For{$i \in\ $DepthFirstSearchPostordering$(X)$}{
			$j \leftarrow $ direct predecessor of $i$ in $X$\\
			$C \leftarrow C :: CNOT_{i, j}$\\
			$P_i \leftarrow P_i \oplus P_j$\\
		}
	}
	\Return $C$
\caption{Parity network synthesis}
\label{code:alltoall}
\end{algorithm}

\begin{figure}[H]
\large
\begin{subfigure}{1\textwidth}
	\begin{minipage}{0.5\textwidth}
	\[
		\begin{tikzpicture} \matrix (m)[ matrix of math nodes, left delimiter  = (, right delimiter = ), ] {
			\bf 0 & 0 & 1 & 1 & 1 & 1 & 1\\
			\bf 1 & 1 & 0 & 0 & 1 & 1 & 1\\
			\bf 1 & 1 & 0 & 1 & 0 & 0 & 1\\
			\bf 0 & 1 & 1 & 1 & 0 & 1 & 1\\
			\bf 1 & 0 & 1 & 1 & 1 & 1 & 1\\
			} ;
		\end{tikzpicture}
	\]
	\end{minipage}
	\begin{minipage}{0.5\textwidth}
	\[
		\begin{tikzpicture}[>=Stealth,->,line width=.7pt] \matrix [matrix of math nodes, column sep={2.0cm,between origins},
		row sep={2.5cm,between origins}, nodes={circle, draw, minimum size=1cm}] {
			& |(2)| 2 &   & |(3)| 3 & \\
			&  & |(5)| 5 & \\
		};
		\draw[transform canvas={xshift=2pt,yshift=-2pt},shorten >= 2pt,shorten <= -2pt] (2) to node[transform canvas={xshift=-2pt}, below]{$-1$} (3);
		\draw[transform canvas={xshift=-2pt,yshift=2pt},shorten >= 2pt,shorten <= -2pt, line width=0.55mm] (3) to node[transform canvas={xshift=2pt}, above]{$-2$} (2);
		\draw[transform canvas={xshift=2pt,yshift=-2pt},shorten >= -1pt] (5) to node[transform canvas={xshift=-2pt, yshift=2pt}, below right]{$0$}  (3);
		\draw[transform canvas={xshift=-2pt,yshift=2pt},shorten >= -1pt] (3) to node[transform canvas={xshift=2pt, yshift=-2pt}, above left]{$-2$} (5);
		\draw[transform canvas={xshift=2pt,yshift=2pt},shorten <= -1pt] (5) to node[transform canvas={xshift=-2pt, yshift=-2pt}, above right]{$-2$} (2);
		\draw[transform canvas={xshift=-2pt,yshift=-2pt},shorten <= -1pt, line width=0.55mm] (2) to node[transform canvas={xshift=2pt, yshift=2pt}, below left]{$-3$} (5);
		\end{tikzpicture}
	\]
	\end{minipage}
	\caption{Parity table and graph $G_{\bs y}$. The minimum weight spanning arborescence of $G_{\bs y}$ is represented by bold arcs.}
\end{subfigure}\\~\\

\begin{subfigure}{1\textwidth}
	\begin{minipage}{0.5\textwidth}
	\[
		\begin{tikzpicture} \matrix (m)[ matrix of math nodes, left delimiter  = (, right delimiter = ), ] {
			\bf 0 & 0 & 1 & 1 & 1 & 1 & 1\\
			\bf 0 & 0 & 0 & 1 & 1 & 1 & 0\\
			\bf 1 & 1 & 0 & 1 & 0 & 0 & 1\\
			\bf 0 & 1 & 1 & 1 & 0 & 1 & 1\\
			\bf 0 & 1 & 1 & 1 & 0 & 0 & 0\\
			} ;
		\end{tikzpicture}
	\]
	\end{minipage}
	\begin{minipage}{0.5\textwidth}
	\[
	\begin{array}{c}
		\Qcircuit @C=1.8em @R=1.4em {
			\lstick{x_1} & \qw & \qw & \qw & \rstick{\!\!\!\!\!\!\!\!\!\!x_1}\\
			\lstick{x_2} & \targ  & \ctrl{1} & \qw & \rstick{\!\!\!\!\!\!\!\!\!\!x_2 \oplus x_5}\\
			\lstick{x_3} & \qw & \targ & \qw  & \rstick{\!\!\!\!\!\!\!\!\!\!x_2 \oplus x_3 \oplus x_4}\\
			\lstick{x_4} & \qw & \qw & \qw & \rstick{\!\!\!\!\!\!\!\!\!\!x_4}\\
			\lstick{x_5} & \ctrl{-3} & \qw & \qw  & \rstick{\!\!\!\!\!\!\!\!\!\!x_5}\\
						 & & & &\\
		}
	\end{array}
	\]
	\end{minipage}
	\caption{State of the parity table after the synthesis of $\bs y$, the additions performed are $P_5 = P_5 \oplus P_2$ and $P_2 = P_2 \oplus P_3$. And circuit corresponding to the synthesis of $\bs y$ with respect to the minimum weight spanning arborescence of $G_{\bs y}$.}
\end{subfigure}
\caption{Example for 1 iteration of Algorithm~\ref{code:alltoall}. The chosen parity $\bs y$ is represented in bold.}
\label{ex:alltoall}
\end{figure}
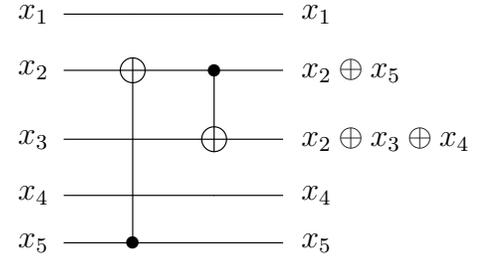~\\

\noindent{\bf Further optimizations.}
The parity $\bs y$ selected in the step $1$ of our algorithm satisfies $\bs y = \operatorname*{argmin} \{h(\bs y) \mid \bs y \in P\}$.
Yet it often arises that multiple parities satisfy this property.
A second selection criteria is needed in order to choose among these parities.
After experimenting with different methods, we found that what gives the best results is to simply choose the parity representing the smallest integer.
This way the parities will be ordered according to their most significant bit as it is done in~\cite{welch}, and the parities will be processed in an order analogous to the Gray code.
As a result and in the same manner as the Gray-Synth algorithm, our algorithm is optimal when the given parity table contains all possible parities.

As previously mentioned, if two sequences $S_i, S_j$ are associated with the same spanning arborescence $X$, then they are equivalent in the sense that $P^{S_i} = P^{S_j}$.
Nonetheless, the circuits produced by $S_i$ and $S_j$ can have different depths.
Thus, by choosing wisely among these equivalent sequences we could enhance the depth performances of our algorithm without affecting the CNOT count of the outputed circuit.
This opportunity is interesting as depth is often considered as a second metric in CNOT circuits synthesis.
Despite that, we decided not to implement such optimization as finding the depth-optimal traversal of $X$ induces a computational overhead.
As a side note, we want to mention that one could also prioritize depth minimization over CNOT count minimization by finding the sequence of additions $S$ that minimizes $h(P^S)$ among all the sequences in $S_{\bs y}$ with a corresponding CNOT circuit of depth $\lceil \log_2 h(\bs y) \rceil$.

\subsection{Benchmarks}
For a parity table $P$ of size $n \times m$ we call density the value $100\times\frac{m}{2^n - 1}$, which is the percentage representation of the ratio between the number of parities in $P$ and the number of possible parities for $n$ qubits.
To evaluate the perfomances of our algorithm we generate random parity tables for 7, 10, 13 and 16 qubits with a density varying from $1\%$ to $100\%$.
We compare our results to the state of the art, namely the Gray-Synth algorithm~\cite{gray} and the Lazy-Synth framework~\cite{lazy}.
Throughout this paper, we have always set the Lazy-Synth depth parameter to $3$ for its recursive search.
Our algorithm as well as the Gray-Synth algorithm have been implemented in Python whereas the Lazy-Synth algorithm has been implemented in C++.
The standard deviation $\sigma$ is not represented in our benchmarks as it is particularly low ($\sigma < 10^{-2}$).

\begin{figure}[H]
    \centering
    \includegraphics[width=0.49\textwidth]{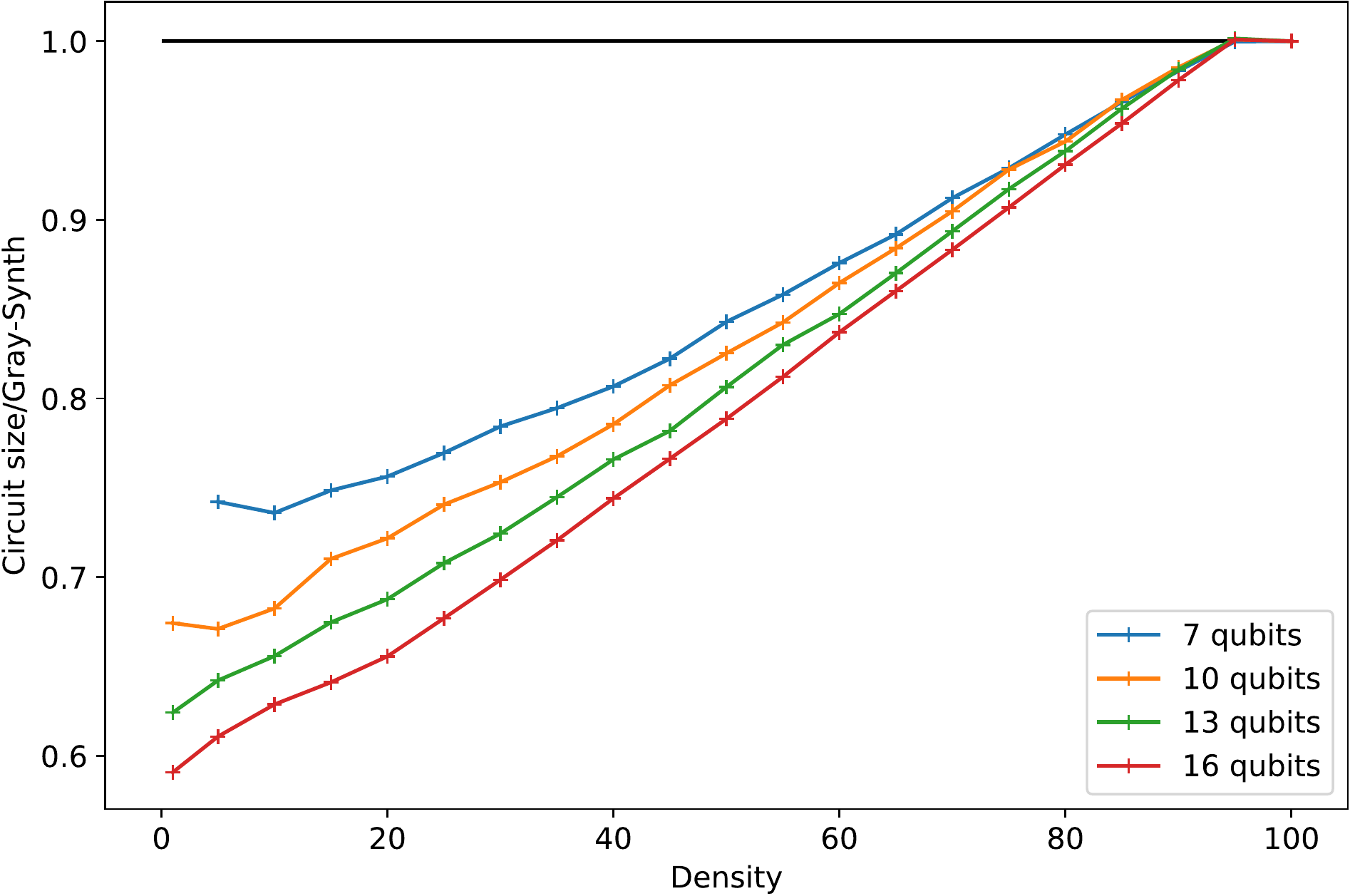}
	\hspace{0.04cm}
    \includegraphics[width=0.49\textwidth]{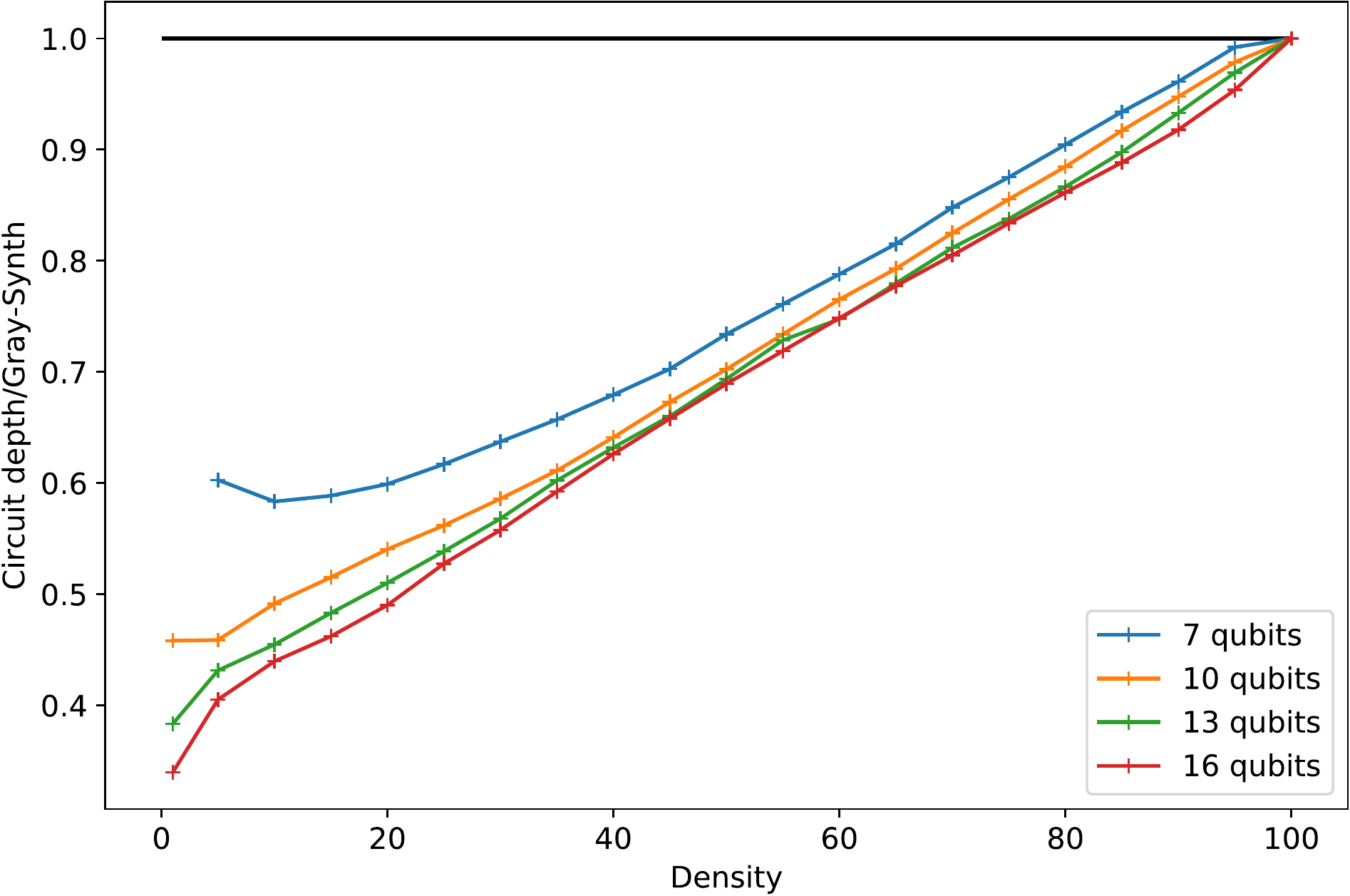}
    \caption{CNOT count and depth of the circuits generated by our algorithm divided by the CNOT count and depth of the circuits outputed by the Gray-Synth algorithm.
    Each point is averaged over $1000, 100, 10$ and $10$ randomly generated parity tables for $7, 10, 13$ and $16$ qubits respectively.}
	\label{fig:cnot_depth_alltoall}
\end{figure}

We first discuss the CNOT count and depth performances of our algorithm in comparison to the Gray-Synth algorithm as presented in Figure~\ref{fig:cnot_depth_alltoall}.
As expected, both algorithms are optimal for full density parity tables.
That being said, our algorithm has a better CNOT depth than the Gray-Synth algorithm for all other densities and it also has a better CNOT count for all parity tables with a density between $1\%$ and $95\%$.
Moreover, we observe that the percentage of CNOT gained over the Gray-Synth algorithm increases as the number of qubits gets larger.
It is therefore reasonable to project that our algorithm outperforms the Gray-Synth algorithm for all phase polynomials acting on larger number of qubits and whose parity table has a density between $1\%$ and $95\%$.

\begin{figure}[H]
    \centering
    \includegraphics[width=0.49\textwidth]{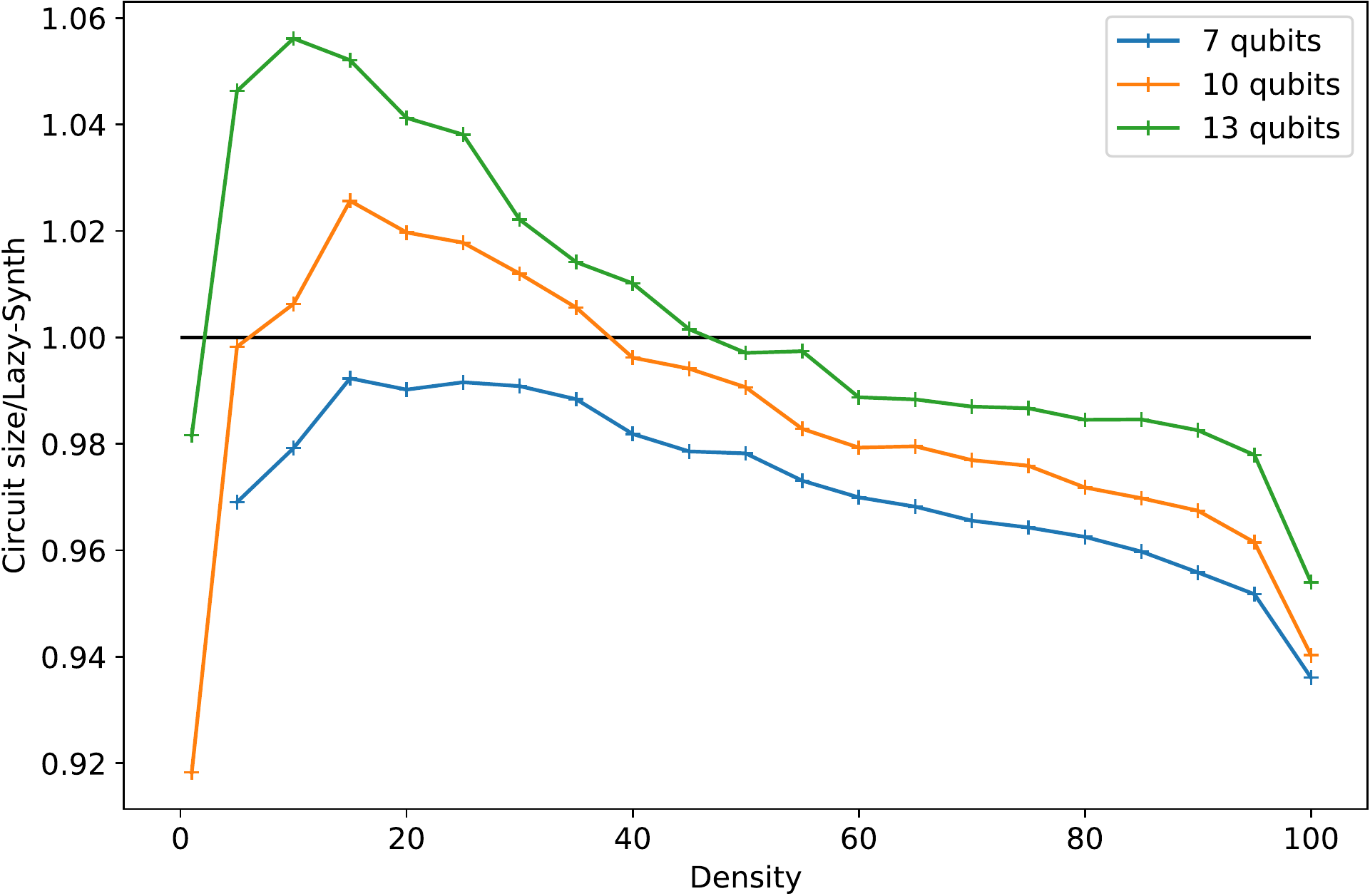}
	\hspace{0.04cm}
    \includegraphics[width=0.49\textwidth]{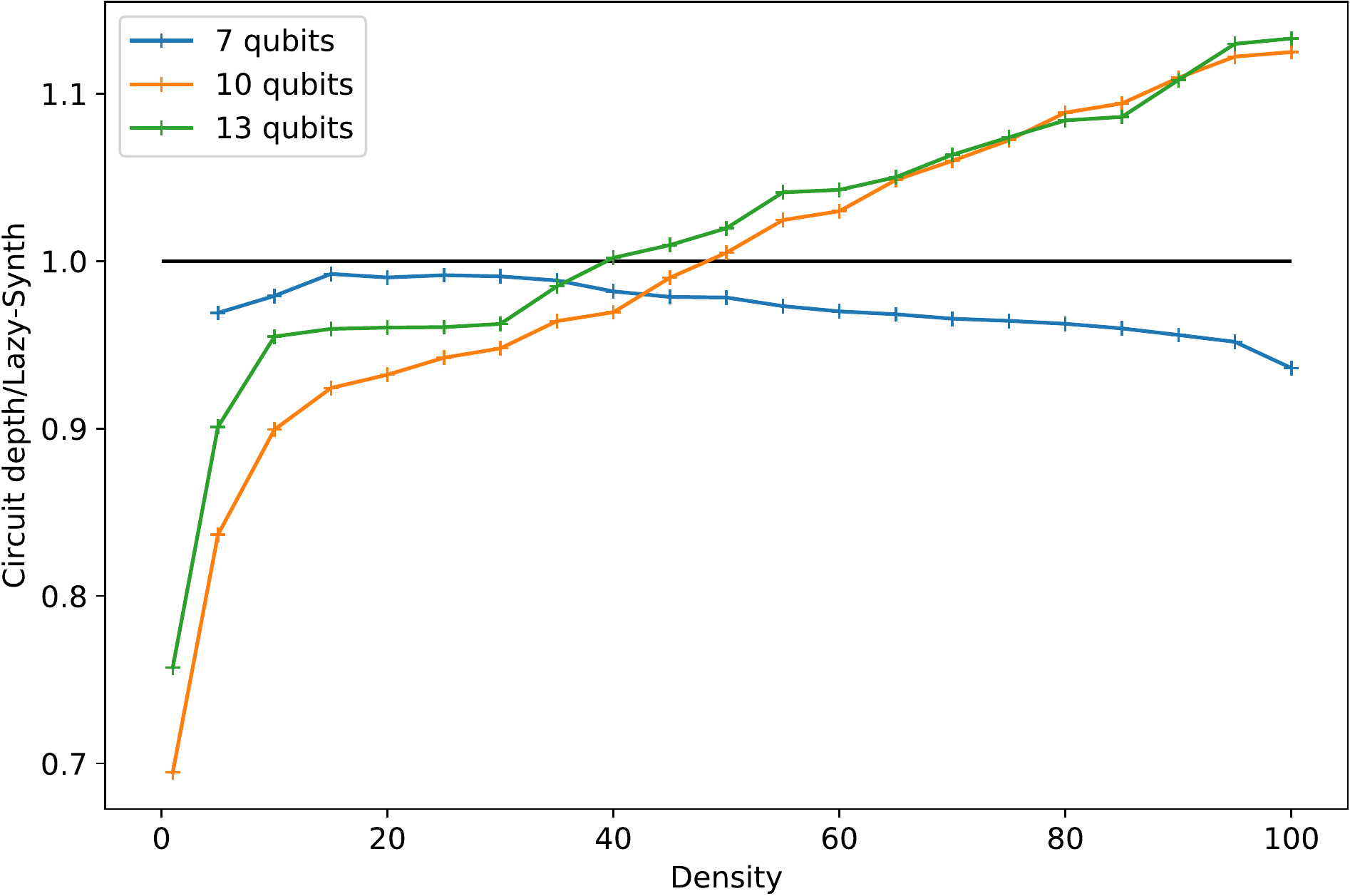}
	\caption{CNOT count and depth of the circuits generated by our algorithm divided by the CNOT count and depth of the circuits outputed by the Lazy-Synth algorithm.
    Each point is averaged over $1000, 100, 10$ and $10$ randomly generated parity tables for $7, 10, 13$ and $16$ qubits respectively.}
	\label{fig:cnot_depth_alltoall_lazy}
\end{figure}

As it can be seen in Figure~\ref{fig:cnot_depth_alltoall_lazy}, for $10$ and $13$ qubits the Lazy-Synth algorithm is solving the parity network synthesis problem with less CNOT than our algorithm for parity tables having a density under $50\%$.
However, this performance gap never exceeds $6\%$ and comes with a tremendous time cost.
The runtimes for the $3$ algorithms are shown in Table~\ref{tab:alltoall_time}, we weren't able to execute the Lazy-Synth algorithm for $16$ qubits due to its important computational time.
As expected by the complexity analysis, our algorithm is slower than the Gray-Synth algorithm but still has a decent enough running time to be applied on large parity tables.
We provide methods to further reduce the running time of our algorithm while preserving good performances in Section~\ref{sec:lowering_complexity}.
Also, the complexity of our algorithm is majored by the creation of the graph $G_{\bs y}$, yet this task can be easily parallelized to significantly reduce the computational time of the algorithm.

\begin{figure}[H]
    \centering
    \includegraphics[width=0.49\textwidth]{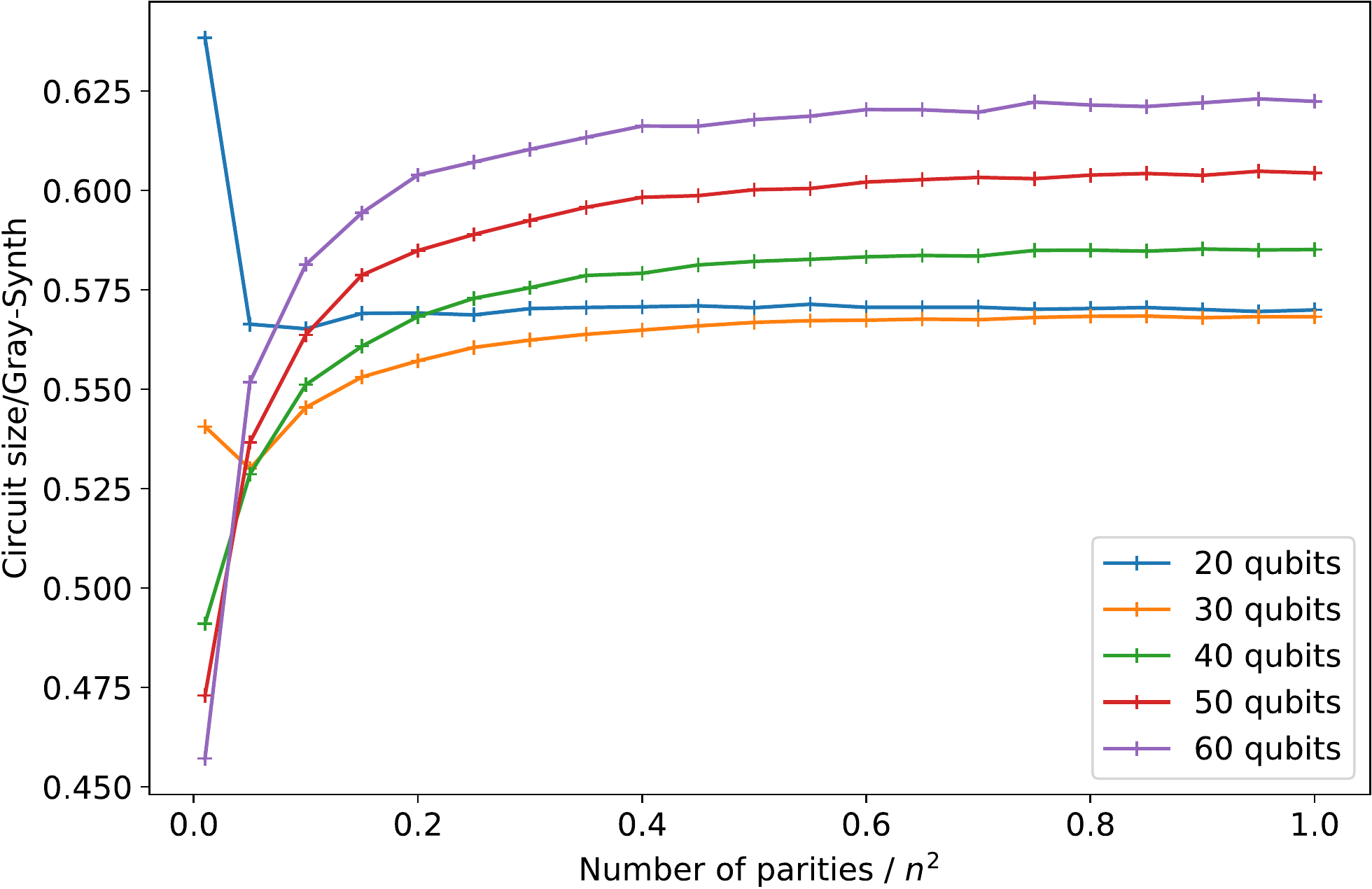}
	\hspace{0.04cm}
    \includegraphics[width=0.49\textwidth]{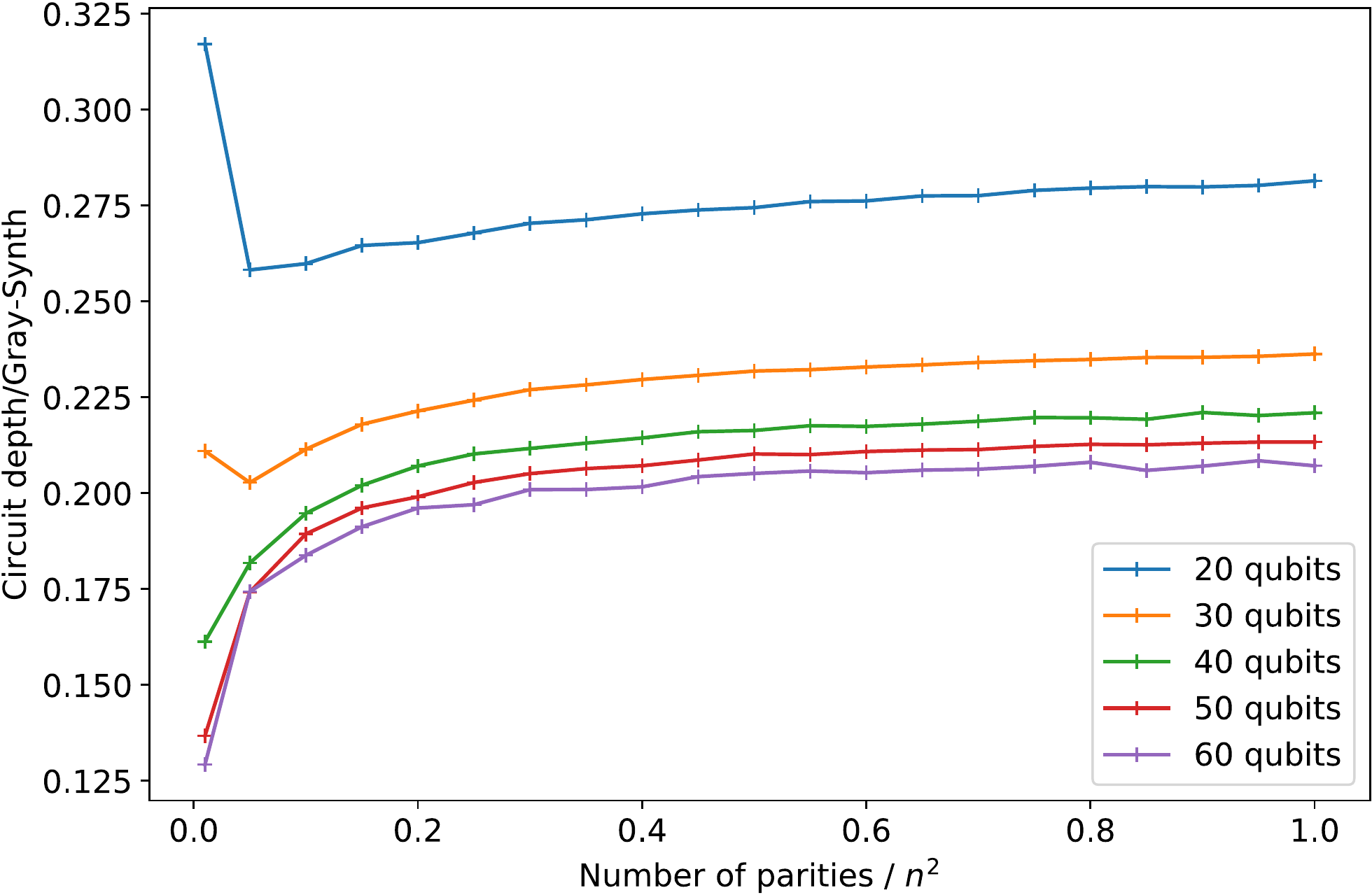}
	\caption{CNOT count and depth of the circuits generated by our algorithm divided by the CNOT count and depth of the circuits outputed by the Gray-Synth algorithm. The number of parities ranges from $1\%$ of $n^2$ to $n^2$ where $n$ is the number of qubits.
	Each point is averaged over $100$ randomly generated parity tables.}
	\label{fig:poly}
\end{figure}

We also performed benchmarks on much lower densities in order to test our algorithm on a higher number of qubits, the results are depicted in Figure~\ref{fig:poly}.
We can see that our algorithm still offers a significant CNOT reduction when compared to the Gray-Synth algorithm for higher number of qubits.
Interestingly, the CNOT count ratio is increasing when the number of qubits increases while the CNOT depth ratio does the opposite.
The computational times for this setting are shown in Table~\ref{tab:alltoall_time_qubits}.
In a similar way as in Table~\ref{tab:alltoall_time}, we can see that the computational time of our algorithm grows faster than the computational time of the Gray-Synth algorithm as the size of the parity table increases.

Overall, with regards to the CNOT metrics, our algorithm offers much better results than the Gray-Synth algorithm and its performances are similar to the Lazy-Synth algorithm but with a viable execution time for large phase polynomials, thus achieving an efficient performance over time ratio.\\

\begin{table}[H]
	\setlength{\tabcolsep}{2pt}
    \caption{Average computational time in seconds of the Gray-Synth algorithm, the proposed algorithm and the Lazy-Synth algorithm.}
	\label{tab:alltoall_time}
	\centerline{
    \scalebox{0.72}{
    \begin{tabular}{|c|ccc|ccc|ccc|cc|}
		\hline
		\multirow{2}{*}{Density} & \multicolumn{3}{c|}{7 qubits} & \multicolumn{3}{c|}{10 qubits} & \multicolumn{3}{c|}{13 qubits} & \multicolumn{2}{c|}{16 qubits}\\
		& Gray-Synth & Proposed & Lazy-Synth & Gray-Synth & Proposed & Lazy-Synth & Gray-Synth & Proposed & Lazy-Synth & Gray-Synth & Proposed \\
		\hline
		1\% & - & - & - &\bf 0.006 & 0.011 & 0.045 &\bf 0.058 & 0.094 & 0.798 &\bf 0.583 & 0.839 \\
		20\% &\bf 0.007 & 0.018 & 0.056 &\bf 0.079 & 0.156 & 1.393 &\bf 0.775 & 1.606 & 94.539 &\bf 8.909 & 37.710 \\
		40\% &\bf 0.011 & 0.032 & 0.113 &\bf 0.111 & 0.303 & 4.283 &\bf 1.167 & 3.708 & 329.136 &\bf 19.042 & 119.998 \\
		60\% &\bf 0.015 & 0.048 & 0.185 &\bf 0.146 & 0.447 & 8.729 &\bf 1.622 & 6.541 & 688.243 &\bf 30.944 & 253.264 \\
		80\% &\bf 0.017 & 0.063 & 0.279 &\bf 0.176 & 0.591 & 14.985 &\bf 2.030 & 10.121 & 1195.277 &\bf 43.895 & 431.766 \\
		100\% &\bf 0.020 & 0.078 & 0.380 &\bf 0.201 & 0.751 & 22.193 &\bf 2.402 & 14.143 & 1748.428 &\bf 56.923 & 652.193 \\
		\hline
	\end{tabular}}}
\end{table}

\begin{table}[H]
    \caption{Average computational time in seconds of the Gray-Synth algorithm and the proposed algorithm for $n^2$ parities where $n$ is the number of qubits.}
	\label{tab:alltoall_time_qubits}
	\centerline{
    \scalebox{1.00}{
    \begin{tabular}{|c|ccccc|}
		\hline
		\multirow{2}{*}{Algorithm} & \multicolumn{5}{c|}{Number of qubits}\\
	   & 20 & 30 & 40 & 50 & 60\\
		\hline
			Gray-Synth & \bf 0.749 & \bf 3.696 & \bf 13.406 & \bf 30.7 & \bf 71.3\\
			Proposed & 0.815 & 4.924 & 22.598 & 81.2 & 243\\
		\hline
	\end{tabular}}}
\end{table}

\section{Parity network synthesis for partial connectivity} 
\label{sec:partialconnectivity}

Many synthesis algorithms are first designed for all-to-all connectivity before being adapted to the case of partial connectivity.
We follow the same process in this work and now demonstrate how our algorithm can be extended to perform architecture-aware synthesis. 

\subsection{Extending the heuristic function}
\label{sec:extending}

As we have seen in Section~\ref{sec:alltoall}, in the case of all-to-all connectivity the cost function $\mathcal{C}: \mathbb{F}_2^n \rightarrow \mathbb{N}$ that gives the minimum CNOT cost for the synthesis of a parity $\bs y$ is equal to $h(\bs y)-1$.
This is utterly incorrect in the case of partial connectivity as the function $h$ doesn't even take into account the connectivity graph of the architecture.
Hence, to extend our algorithm for constrained architectures we first need to determine the value of our heuristic function $\mathcal{C}$ for the case of partial connectivity.
For this purpose we will rely on Steiner trees, which are commonly used when it comes to designing architecture-aware synthesis algorithms~\cite{griend, lazy, mosca, nash, staq}.\\

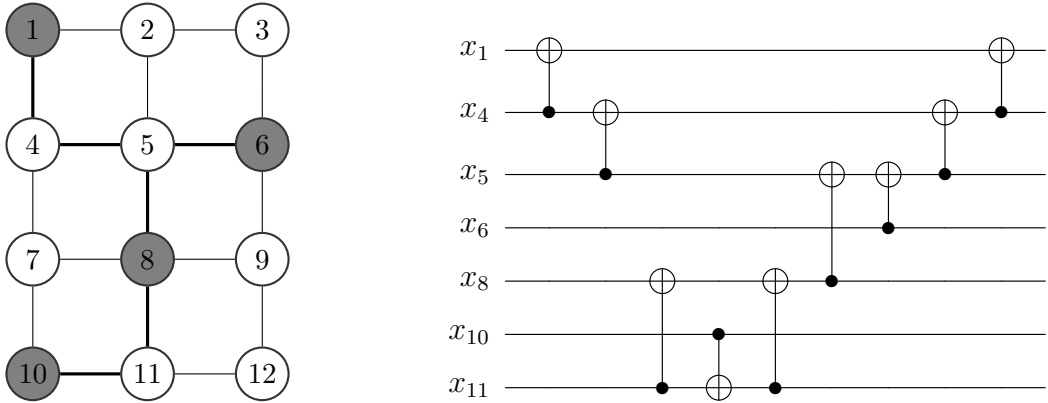
\begin{figure}[H]
\large
	\begin{minipage}{0.5\textwidth}
	\[
		\begin{tikzpicture}[scale=1.22, anchor=west]
			\tikzstyle{w}=[circle,draw=black!80, thick, font=\normalsize, minimum size=0.7cm, inner sep=0pt]
			\tikzstyle{b}=[circle,fill=gray,draw=black!80, thick, font=\normalsize, minimum size=0.7cm, inner sep=0pt]

			\node[b] (v1) at (0,0) {1};
			\node[w] (v2) at (1.25,0) {2};
			\node[w] (v3) at (2.5,0) {3};
			\node[w] (v4) at (0,-1.25) {4};
			\node[w] (v5) at (1.25,-1.25) {5};
			\node[b] (v6) at (2.5,-1.25) {6};
			\node[w] (v7) at (0,-2.5) {7};
			\node[b] (v8) at (1.25,-2.5) {8};
			\node[w] (v9) at (2.5,-2.5) {9};
			\node[b] (v10) at (0,-3.75) {10};
			\node[w] (v11) at (1.25,-3.75) {11};
			\node[w] (v12) at (2.5,-3.75) {12};

			\draw[very thick] (v1) -- (v4);
			\draw[very thick] (v4) -- (v5);
			\draw[very thick] (v5) -- (v6);
			\draw[very thick] (v5) -- (v8);
			\draw[very thick] (v8) -- (v11);
			\draw[very thick] (v10) -- (v11);
			\foreach \from/\to in {v1/v2, v2/v3, v4/v5, v5/v6, v7/v8, v8/v9, v2/v5, v3/v6, v4/v7, v5/v8, v6/v9, v7/v10, v11/v12, v12/v9}
				\draw (\from) -- (\to);
		\end{tikzpicture}
	\]
	\end{minipage}
	\begin{minipage}{0.5\textwidth}
	\[
	\begin{array}{c}
		\Qcircuit @C=1.0em @R=1.18em {
			\lstick{x_1}    & \targ & \qw & \qw & \qw & \qw & \qw & \qw & \qw & \targ & \qw \\
			\lstick{x_4}    & \ctrl{-1} & \targ & \qw & \qw & \qw & \qw & \qw & \targ & \ctrl{-1} & \qw \\
			\lstick{x_5}    & \qw & \ctrl{-1} & \qw & \qw & \qw & \targ & \targ & \ctrl{-1} & \qw & \qw \\
			\lstick{x_6}    & \qw & \qw & \qw & \qw & \qw & \qw & \ctrl{-1} & \qw & \qw & \qw \\
			\lstick{x_8}    & \qw & \qw & \targ & \qw & \targ & \ctrl{-2} & \qw & \qw & \qw & \qw \\
			\lstick{x_{10}} & \qw & \qw & \qw & \ctrl{1} & \qw & \qw & \qw & \qw & \qw & \qw \\
			\lstick{x_{11}} & \qw & \qw & \ctrl{-2} & \targ & \ctrl{-2} & \qw & \qw & \qw & \qw & \qw \\
		}
	\end{array}
	\]
	\end{minipage}
	\caption{Synthesis example for the parity $x_1 \oplus x_6 \oplus x_8 \oplus x_{10}$ on a grid architecture. The gray nodes are the terminals and the Steiner tree is represented by bold edges. The chosen root for the spanning arborescence is the vertex $1$.}
\label{ex:steiner}
\end{figure}

\noindent{\bf Steiner trees.}
Given a graph $G=(V, E)$ and a set of vertices $S \subseteq V$, the Steiner tree problem consists of finding the minimum tree $T=(V_T, E_T)$, called Steiner tree, such that $T$ is a subgraph of $G$ and $S \subseteq V_T$.
The vertices in $S$ are called terminals and the vertices in $V_T \setminus S$ are called Steiner nodes.
To perform the synthesis of a parity $\bs y$, a Steiner tree is required since we must interconnect the vertices in $S=\{i \mid y_i = 1\}$.
Once the Steiner tree $T$ is computed, two steps must be carried out:
\begin{enumerate}
	\item Fill-in all Steiner nodes such that for all $u \in V_T$ we have $y_u = 1$.
	\item Perform the sequence of additions associated with a successors-first traversal of any spanning arborescence of $T$.
\end{enumerate}
An example of this process is given Figure~\ref{ex:steiner}, and the pseudo-code to perform the fill-in step is provided in Algorithm~\ref{code:fillin}. 
We use the notation $G[X]$ to refer to the induced subgraph of $G$ formed by the subset of vertices $X$.
For the second step it is exactly the same procedure as the for loop in Algorithm~\ref{code:alltoall} for an arbitrary spanning arborescence of $T$.
The fill-in step requires $|V_T \setminus S|$ additions and the second step takes $|V_T| - 1$ additions, hence the total CNOT cost for the synthesis of $\bs y$ is $\mathcal{C}(\bs y) = |V_T \setminus S| + |V_T| - 1 = 2|V_T| - |S| - 1$.
Solving optimally the Steiner tree problem would minimize $C(\bs y)$, unfortunately finding the minimal Steiner tree is NP-hard~\cite{karp}.
We will therefore rely on an approximation algorithm proposed by Takashashi et al.~\cite{takahashi} that has an approximation ratio of $2 - 2/|S|$.
The algorithm is presented in pseudo-code in Algorithm~\ref{code:steinertree}, it starts by adding an arbitrary vertex of $S$ to the Steiner tree $T$ and then constructs $T$ by iteratively adding the shortest path between $T$ and one vertex of $S$ not yet in $T$.
Its runtime is $\mathcal{O}(|S||V|^2)$, although it can be lowered to $\mathcal{O}(|V|^2)$ if all the shortest paths are provided.\\

\begin{algorithm}[H]
	\SetAlgoLined
	\SetArgSty{textnormal}
	\SetKwFunction{proc}{FillIn}
	\SetKwProg{Fn}{procedure}{}{}
	\Fn{\proc{$T$, $S$}}{
		$C \leftarrow$ new empty circuit\\
		$F \leftarrow T[V_T \setminus S]$\\
		\While{$F$ \text{non-empty}}{
			$u \leftarrow$ a leaf of $F$\\
			$v \leftarrow $ any vertex in $S \cap T.neighbors(u)$\\
			$C \leftarrow C :: CNOT_{u, v}$\\
			$S \leftarrow S \cup \{u\}$\\
			$F \leftarrow F[V_{F} \setminus \{u\}]$\\
		}
		\Return $C$
	}
\caption{Arbitrary fill-in}
\label{code:fillin}
\end{algorithm}~\\

\begin{algorithm}[H]
	\SetAlgoLined
	\SetArgSty{textnormal}
	\SetKwFunction{proc}{SteinerTree}
	\SetKwProg{Fn}{procedure}{}{}
	\Fn{\proc{$S$, $paths$}}{
		$u \leftarrow $ any vertex in $S$\\
		$S \leftarrow S \setminus \{u\}$\\
		$T \leftarrow $ Graph($\{u\}, \emptyset$)\\
		\While{$S$ \text{non-empty}}{
			$\displaystyle u, v \leftarrow \operatorname*{argmin}_{u,v} \{|paths_{u,v}| \mid u \in T, v \in S\}$\\
			$T \leftarrow T \cup paths_{u, v}$\\
			$S \leftarrow S \setminus \{v\}$\\
		}
		\Return $T$
	}
\caption{Steiner tree}
\label{code:steinertree}
\end{algorithm}

\subsection{Architecture-aware algorithm}
In this section we present an architecture-aware version of Algorithm~\ref{code:alltoall} by relying on the heuristic function $\mathcal{C}$ defined in Section~\ref{sec:extending}.
The pseudo-code of the algorithm is provided in Algorithm~\ref{code:architecture}.
We reuse some notations of Section~\ref{sec:alltoall}: $P$ is a parity table, $\bs y$ is a parity of $P$ and $G_{\bs y}$ is the parity graph associated with $\bs y$.
We define $P^X$ as being the state of the parity table $P$ after performing the sequence of additions associated with any successors-first traversal of the arborescence $X$.
We denote by $T_{\bs y}$ the Steiner tree of $\bs y$ where the terminals are the vertices in the set $S=\{i \mid y_i = 1\}$.

We consider again the two steps process described in Section~\ref{sec:alltoall} and modify it to take into account the architecture's connectivity.
Namely, we choose the parity $\bs y$ that minimizes $\mathcal{C}(\bs y)$ and we perform the synthesis of $\bs y$ in a architecture-aware manner such as described in Section~\ref{sec:extending}.
In most cases, the number of minimum size fill-in of $T_{\bs y}$ is exponential with respect to the number of terminals.
As we want our algorithm to be scalable we get over this step by doing an arbitrary fill-in as presented in Algorithm~\ref{code:fillin}.
Then, we must perform the sequence of additions associated with a successors-first traversal of any spanning arborescence $X$ of $T_{\bs y}$.
Our heuristic function $\mathcal{C}$ isn't merely based on the Hamming weight anymore, it implies that we cannot rely again on the minimum weight spanning arborescence to choose among the spanning arborescences of $T_{\bs y}$ as we have done in Section~\ref{sec:alltoall}.
However, since $T_{\bs y}$ is a tree it only has $|V_{T_{\bs y}}|$ different spanning arborescences (a tree can have exactly one spanning arborescence rooted at each of its vertices), we can compute all the possibilities to choose the one that gives the best results and still have a polynomial time algorithm.

Let $\bs c^X = \text{sort}\{\mathcal{C}(\bs y) \mid \bs y \in P^X\}$ be the cost vector sorted in ascending order with respect to a spanning arborescence $X$ of $T_{\bs y}$.
As we don't rely on the minimum weight spanning arborescence anymore, we are not compelled to choose the spanning arborescence $X$ of $T_{\bs y}$ that minimizes $\sum_{\bs y \in P^X} \mathcal{C}(\bs y)$.
Our experiments have shown that choosing the spanning arborescence $X$ of $T_{\bs y}$ such that $\bs c^X_1$ is minimal leads to better results.
If several arborescences satisfy this property then we choose among them by taking the spanning arborescence $X$ such that $\bs c^X_2$ is minimal.
We repeat the process for $\bs c^X_i$ where $i \in [3, \ldots, m]$ until we only have one spanning arborescence left or until the end of the cost vectors is reached.
The chosen spanning arborescence $X$ of $T_{\bs y}$ then satisfies 
\begin{align*}
X &= \operatorname*{argmin}\{\bs c^X \mid X \text{ is a spanning arborescence of }T_{\bs y}\}\\
  &= \operatorname*{argmin}\{\text{sort}\{\mathcal{C}(\bs y) \mid \bs y \in P^X\} \mid X \text{ is a spanning arborescence of }T_{\bs y}\}\}.
\end{align*}

\begin{algorithm}[H]
	\SetAlgoLined
	\SetArgSty{textnormal}
	\SetKwInput{KwInput}{Input}
	\SetKwInput{KwOutput}{Output}
	\KwInput{A parity table $P$ and a connectivity graph $G$}
	\KwOutput{Circuit synthesizing a parity network associated with $P$ and compliant with the architectural constraints described by $G$}
	$C \leftarrow$ new empty circuit\\
	$paths \leftarrow $ShortestPaths$(G)$\\
	\While{$P$ \text{non-empty}}{
		$\displaystyle \bs y \leftarrow \operatorname*{argmin}_{\bs y} \{\mathcal{C}(\bs y) \mid \bs y \in P\}$\\
		$P \leftarrow P \setminus \bs y$\\
		$S \leftarrow \{i \mid y_i=1\}$\\
		$T \leftarrow$ SteinerTree($S$, $paths$)\\

		\For{$CNOT_{i,j} \in\ $FillIn($T$, $S$)}{
			$P_i \leftarrow P_i \oplus P_j$\\
			$C \leftarrow C :: CNOT_{i,j}$\\
		}

		$\displaystyle X \leftarrow \operatorname*{argmin}_X \{$sort$\{\mathcal{C}(\bs y) \mid \bs y \in P^X\} \mid X$ is a spanning arborescence of $T\}$\\
		\For{$i \in\ $DepthFirstSearchPostordering$(X)$}{
			$j \leftarrow $ direct predecessor of $i$ in $X$\\
			$C \leftarrow C :: CNOT_{i, j}$\\
			$P_i \leftarrow P_i \oplus P_j$\\
		}
	}
	\Return $C$
\caption{Architecture-aware parity network synthesis}
\label{code:architecture}
\end{algorithm}~\\

\noindent{\bf Complexity analysis.}
Let $n$ be the number of qubits and $m$ the number of parities.
Algorithm~\ref{code:steinertree} compute the Steiner tree of a parity with a complexity of $\mathcal{O}(n^2)$.
The complexity of Algorithm~\ref{code:architecture} is majored by the task of finding the optimal spanning arborescence $X$, which takes $\mathcal{O}(mn^3)$ operations as it requires to compute $\mathcal{O}(mn)$ Steiner trees.
The algorithm performs $m$ iterations so the overall complexity is $\mathcal{O}(m^2n^3)$.
Some methods to further reduce the complexity of our algorithm are described in Section~\ref{sec:lowering_complexity}.\\

\noindent{\bf Further optimizations.}
We refer to the implementation cost of a Steiner tree $T_{\bs y}$ as the value $\mathcal{C}(\bs y)$.
In most cases, in order to find the spanning arborescence $X$ satisfying $X = \operatorname*{argmin}\{\text{sort}\{\mathcal{C}(\bs y) \mid \bs y \in P^X\}\} = \operatorname*{argmin}\{\bs c^X\}$, we don't actually need to know all the values of the vector $\bs c^X$.
Our experiments show that the CNOT performances of our algorithm doesn't change when we only consider the $K=10$ first values of $\bs c^X$.
Indeed, it is rather rare to have $2$ arborescences $X_1$, $X_2$ such that $\bs c^{X_1}_i = \bs c^{X_2}_i \  \forall i \in [1, \ldots, K=10]$.
Also, when it happens, then choosing $X_1$ or $X_2$ doesn't make a significant difference in the algorithm performances.
Consequently, to determine the values $\bs c^{X}_i \ \forall i \in [1, \ldots, K]$ for an arborescence $X$, we only have to compute the $K$ Steiner trees of $\{T_{\bs y} \mid \bs y \in P^X\}$ having a minimum implementation cost $\mathcal{C}(\bs y)$.

In order to find the $K$ minimum cost Steiner trees we propose an algorithm that simultaneously constructs all the Steiner trees and stops when the $K$ minimum cost Steiner trees are found.
In this algorithm, only $K$ Steiner trees will be completely computed and all the others will be partially constructed, thus saving a considerable amount of time.
The pseudo-code of the algorithm is given in Algorithm~\ref{code:kminimum}, it takes as input the shortest paths of the graph, an integer $K$ and a set $S$ of sets $S_1, \ldots, S_m$ of terminals from which we want to find the $K$ minimum cost Steiner trees.\\

\begin{algorithm}[H]
	\SetAlgoLined
	\SetArgSty{textnormal}
	\SetKwFunction{proc}{MinimumCostSteinerTrees}
	\SetKwProg{Fn}{procedure}{}{}
	\Fn{\proc{$S$, $paths$, $K$}}{
		$R \leftarrow$ new empty list\\
		$L \leftarrow$ new list of empty stacks\\
		$cost \leftarrow 0$\\
		\For{$i \in \{1, \ldots, |S|\}$}{
			$u \leftarrow $ any vertex in $S_i$\\
			$S_i \leftarrow S_i \setminus \{u\}$\\
			$T_i \leftarrow $ Graph($\{u\}, \emptyset$)\\
			$L_0.$push$(i)$\\
		}
		\While{$L$ \text{non-empty}}{
			$Q \leftarrow L$.popleft\\
			\While{$Q$ \text{non-empty}}{
				$i \leftarrow Q.$pop\\
				\If{$S_i$ is empty}{
					$R \leftarrow R :: (cost,\; T_i)$\\
					\If{$|R| = K$}{
						\Return $R$
					}
				}
				$\displaystyle u, v \leftarrow \operatorname*{argmin}_{u,v} \{|paths_{u,v}| \mid u \in T_i, v \in S_i\}$\\
				$T_i \leftarrow T_i \cup paths_{u, v}$\\
				$S_i \leftarrow S_i \setminus \{v\}$\\
				$j \leftarrow 1 + 2 \times (|paths_{u, v}|-2)$\\
				$L_{j}.$push$(i)$\\
			}
			$cost \leftarrow cost + 1$\\
		}
		\Return $R$
	}
\caption{$K$ minimum cost Steiner trees}
\label{code:kminimum}
\end{algorithm}

\subsection{Lowering the complexity} \label{sec:lowering_complexity}
In this section we describe some methods that can be applied to our algorithms in order to reduce their complexity and achieve the best trade-off possible between performances and running time constraints.
We demonstrate these methods in the case of constrained architectures, although they can also be applied in the same way on Algorithm~\ref{code:alltoall} for all-to-all connectivity.\\

\noindent{\bf Greedy method.}
For this method we consider a greedier version of Algorithm~\ref{code:architecture} by choosing an arbitrary spanning arborescence instead of the one satisfying $X = \operatorname*{argmin} \{\text{sort}\{\mathcal{C}(\bs y) \mid \bs y \in P^X\}$.
The algorithm then simply consists in choosing the parity $\bs y$ minimizing $\mathcal{C}(\bs y)$, performing the synthesis of $\bs y$ optimally and reiterating until all parities have been synthesized.
The complexity of this algorithm is $\mathcal{O}(m^2n^2)$.\\

\noindent{\bf Sliding window.}
Another way of lowering the complexity of our algorithm is to only consider the parities that are in the range of a sliding window of size $\alpha$ that runs over the parity table.
At each iteration of our algorithm the parity which has just been synthesized is removed from the window and another parity not yet considered is added to the window.
With this method the complexity of our algorithm is lowered to $\mathcal{O}(\alpha mn^3)$.
By combining the greedy and sliding window methods, the complexity of the algorithm is further lowered to $\mathcal{O}(\alpha mn^2)$.

\subsection{Benchmarks}
We compare our algorithms with the algorithm proposed by Meijer-van de Griend et al.~\cite{duncan}, the phase polynomials synthesis algorithm implemented in Staq~\cite{staq} and the Lazy-Synth algorithm~\cite{lazy}.
Beside Algorithm~\ref{code:architecture}, we benchmark 3 modified versions of it stemming from Section~\ref{sec:lowering_complexity}: with a sliding window of size $50$, with the greedy method and with both the greedy method and a sliding window of size $50$.
All algorithms have been implemented in Python except for the Lazy-Synth algorithm which has been implemented in C++.
The standard deviation $\sigma$ is not represented in our benchmarks as it is particularly low ($\sigma < 10^{-2}$).

\begin{figure}[H]
\centering
	\begin{subfigure}{1\textwidth}
		\includegraphics[height=0.29\textwidth]{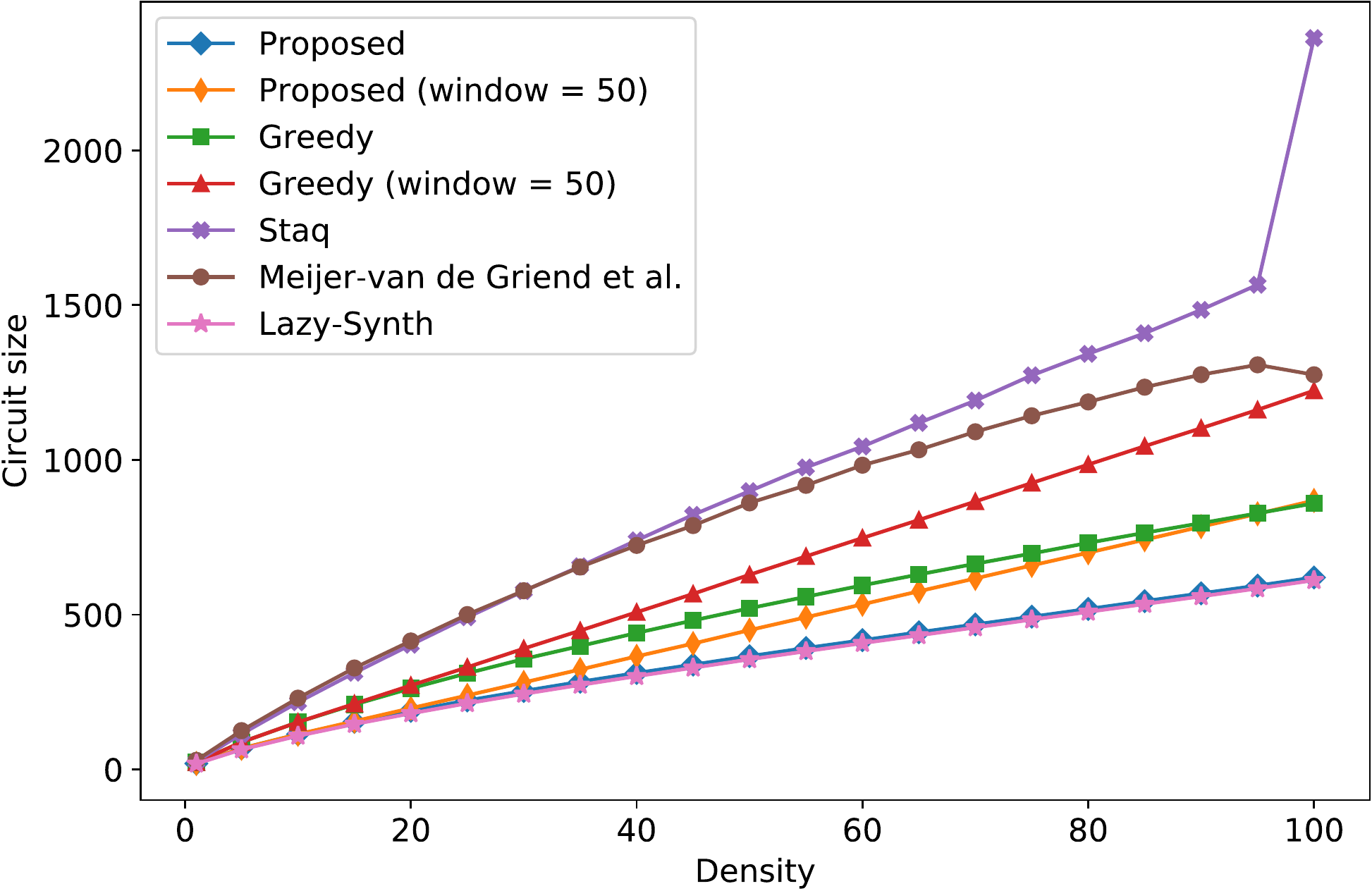}
		\hspace{0.2cm}
		\includegraphics[height=0.29\textwidth]{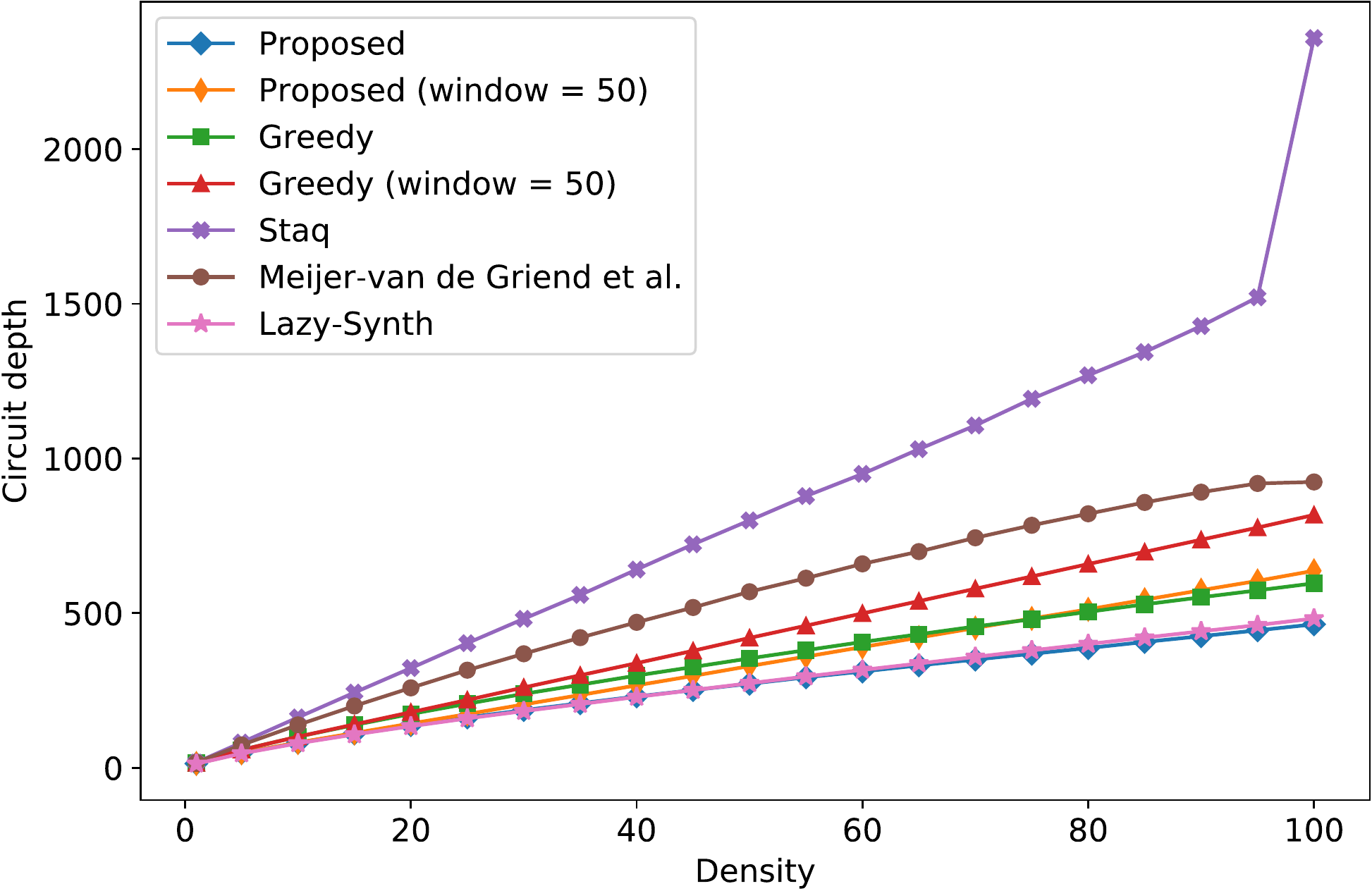}
		\vspace{0.4cm}\\
		\centering
		\includegraphics[height=0.29\textwidth]{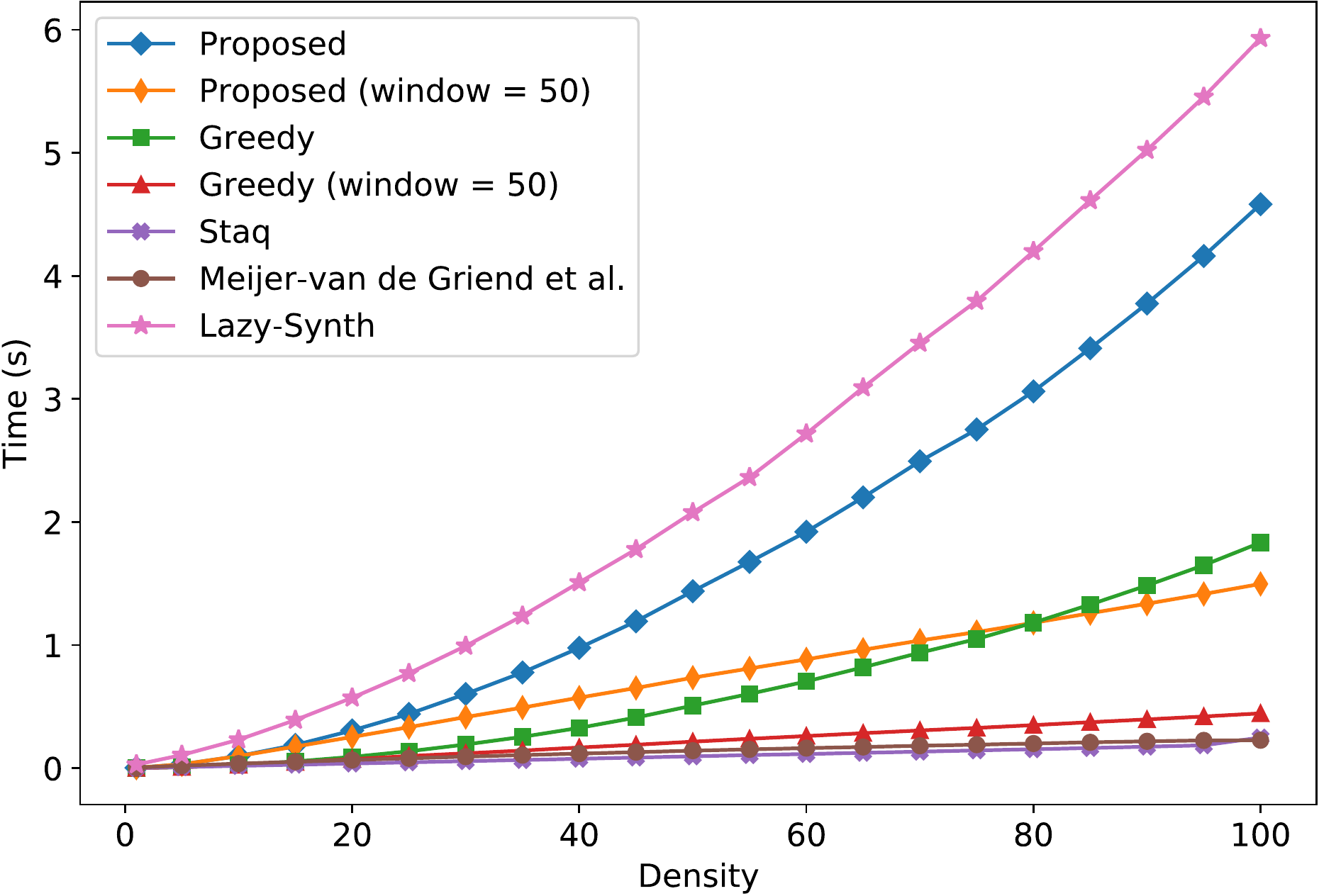}
		\caption{Grid $3 \times 3$}
		\label{fig:grid}
	\end{subfigure}
	\vspace{0.5cm}

	\begin{subfigure}{1\textwidth}
		\includegraphics[height=0.29\textwidth]{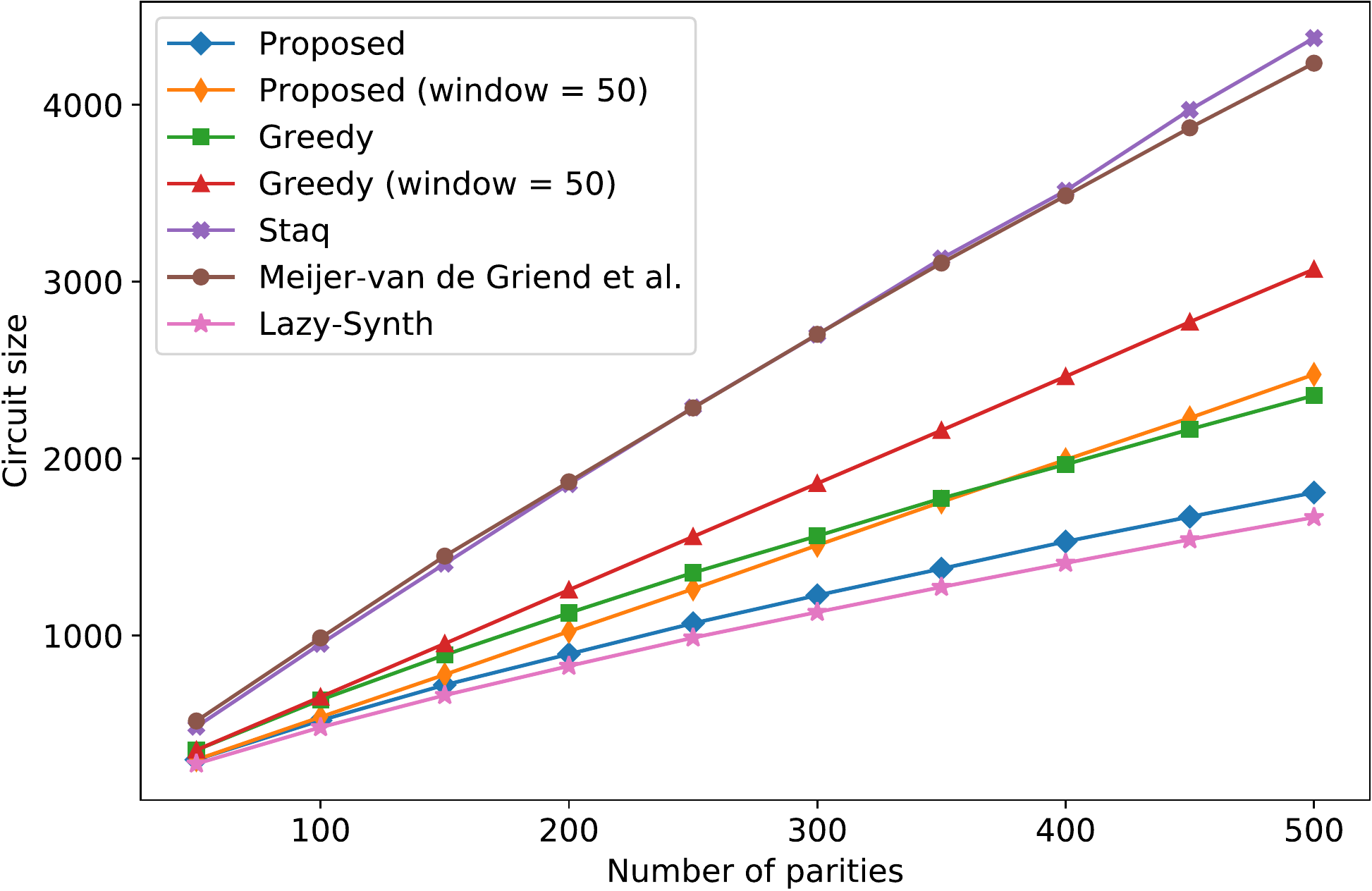}
		\hspace{0.2cm}
		\includegraphics[height=0.29\textwidth]{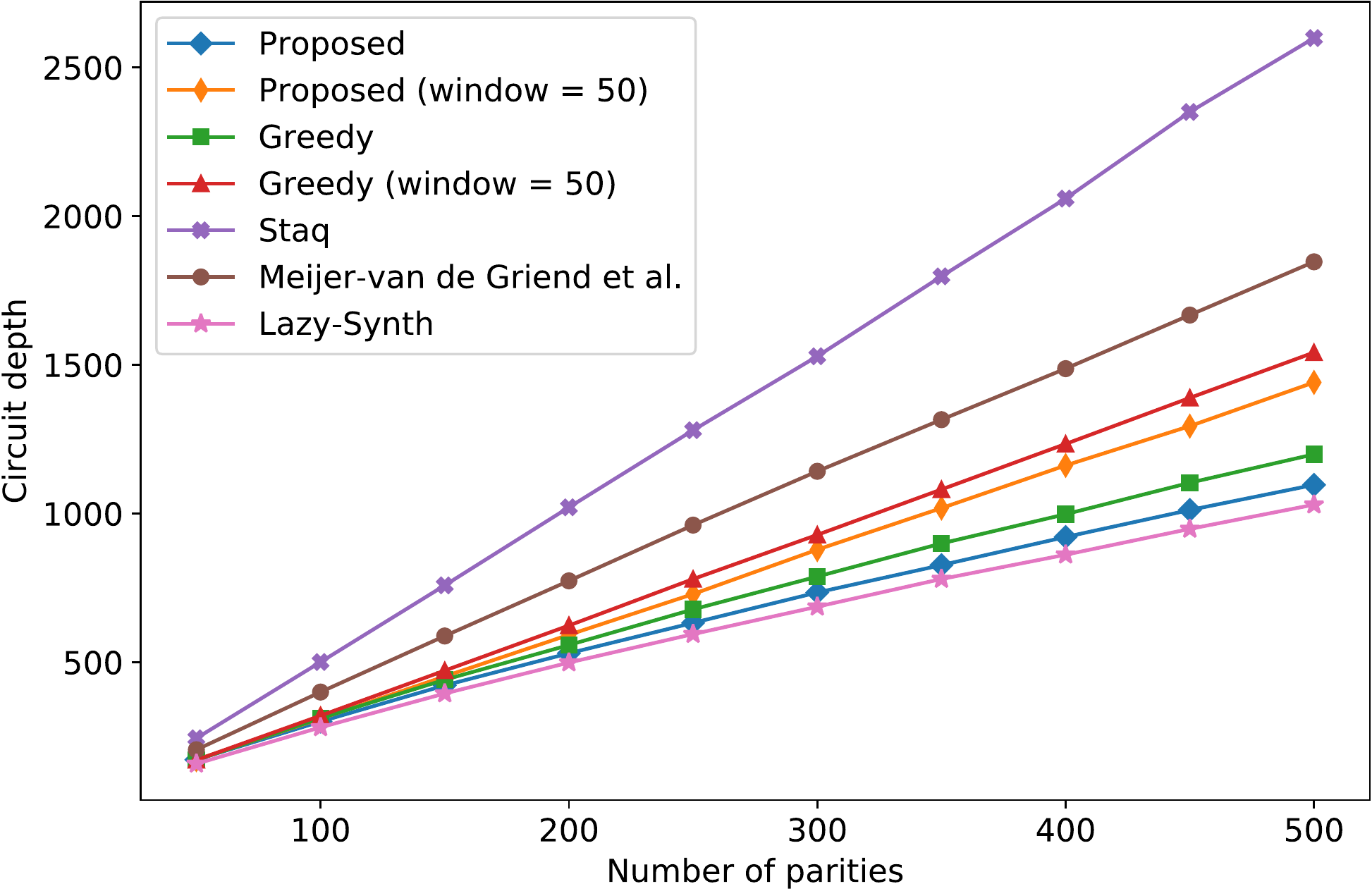}
		\vspace{0.4cm}\\
		\centering
		\includegraphics[height=0.29\textwidth]{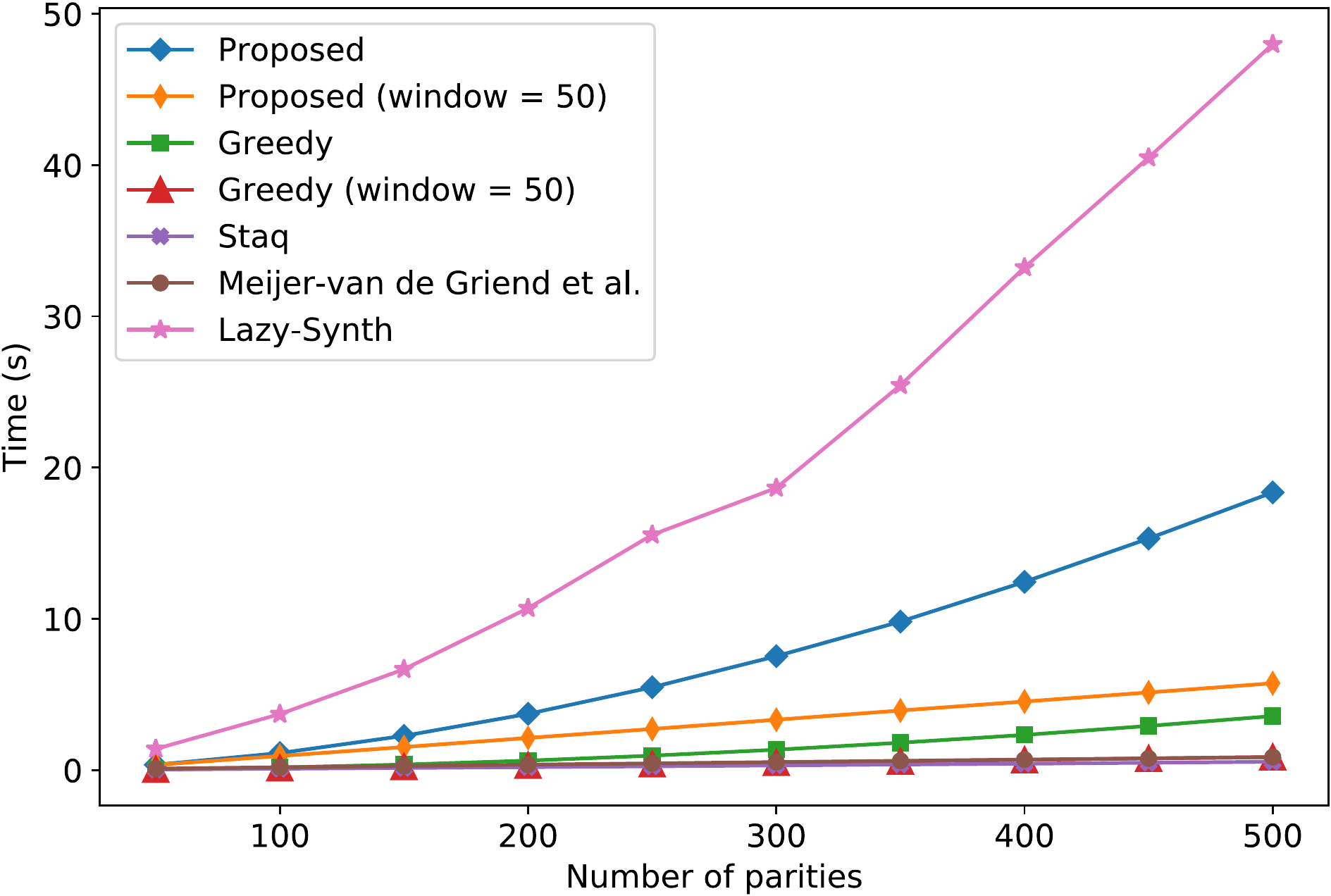}
		\caption{14 qubits IBM Melbourne}
		\label{fig:melbourne}
	\end{subfigure}
	
    \caption{Scaling of the circuit size, circuit depth and computational time on a $3 \times 3$ grid architecture and the 14 qubits IBM's Melbourne architecture.
    Each point is averaged over $1000$ randomly generated parity tables.}
	\label{fig:architecture_bench}
\end{figure}

Figure~\ref{fig:grid} compares the algorithms on a $3 \times 3$ grid with a density ranging from $1\%$ to $100\%$.
In this setting, our algorithm and the Lazy-Synth algorithm both offer the best CNOT performances, our algorithm being faster.
In Figure~\ref{fig:melbourne}, for the Melbourne architecture, the Lazy-Synth algorithm provides the best performances with regards to the CNOT metrics but has an important running time.
All the variants of Algorithm~\ref{code:architecture} have a better CNOT count and CNOT depth than the algorithms proposed by Meijer-van de Griend et al. and Staq.
Morever, the greedy version with a sliding window of size $50$ has a similar running time.
We tested the algorithms on various architectures and didn't notice any significant changes.

\begin{figure}[H]
	\includegraphics[width=0.49\textwidth]{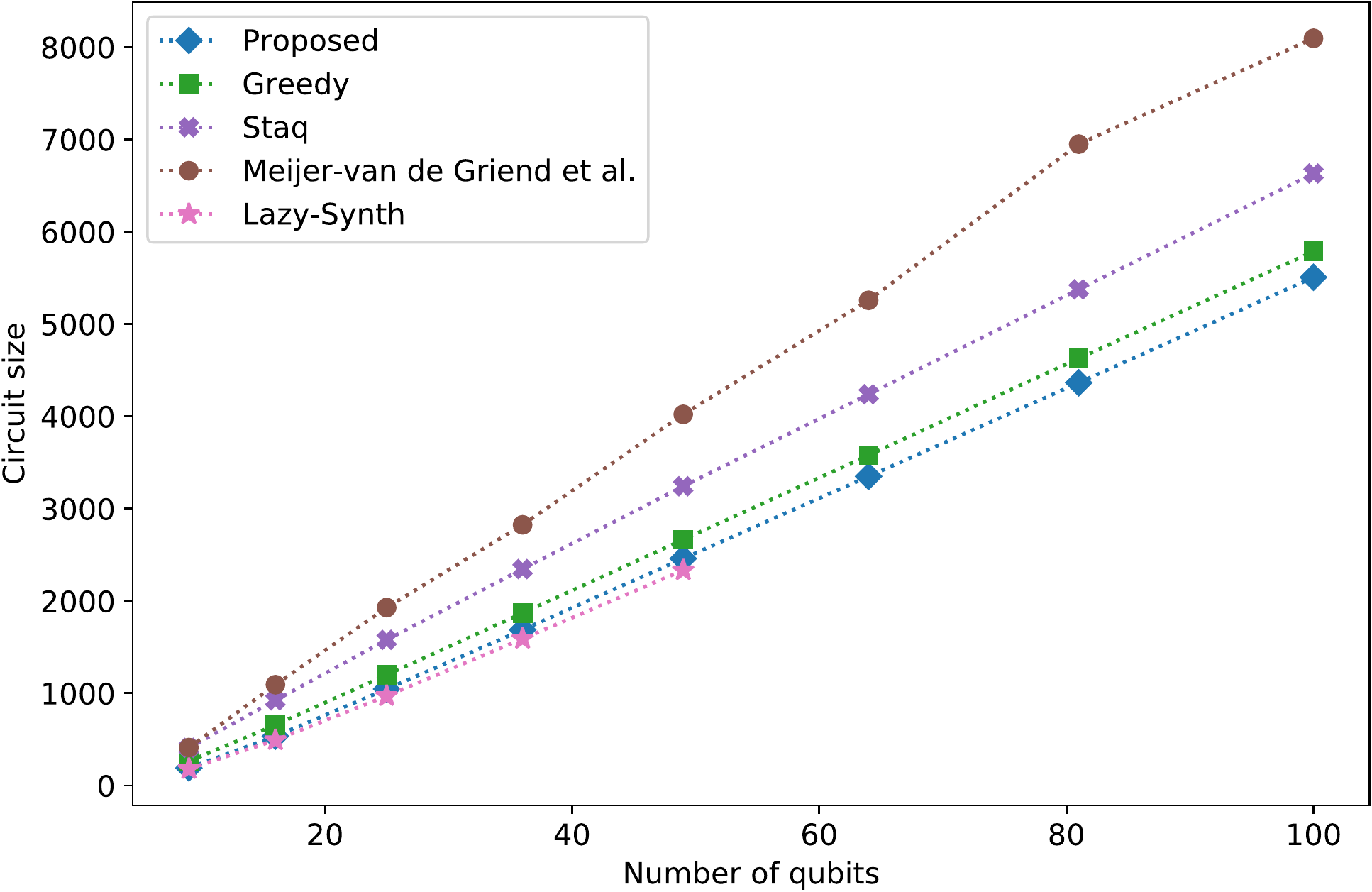}
	\hspace{0.04cm}
	\includegraphics[width=0.49\textwidth]{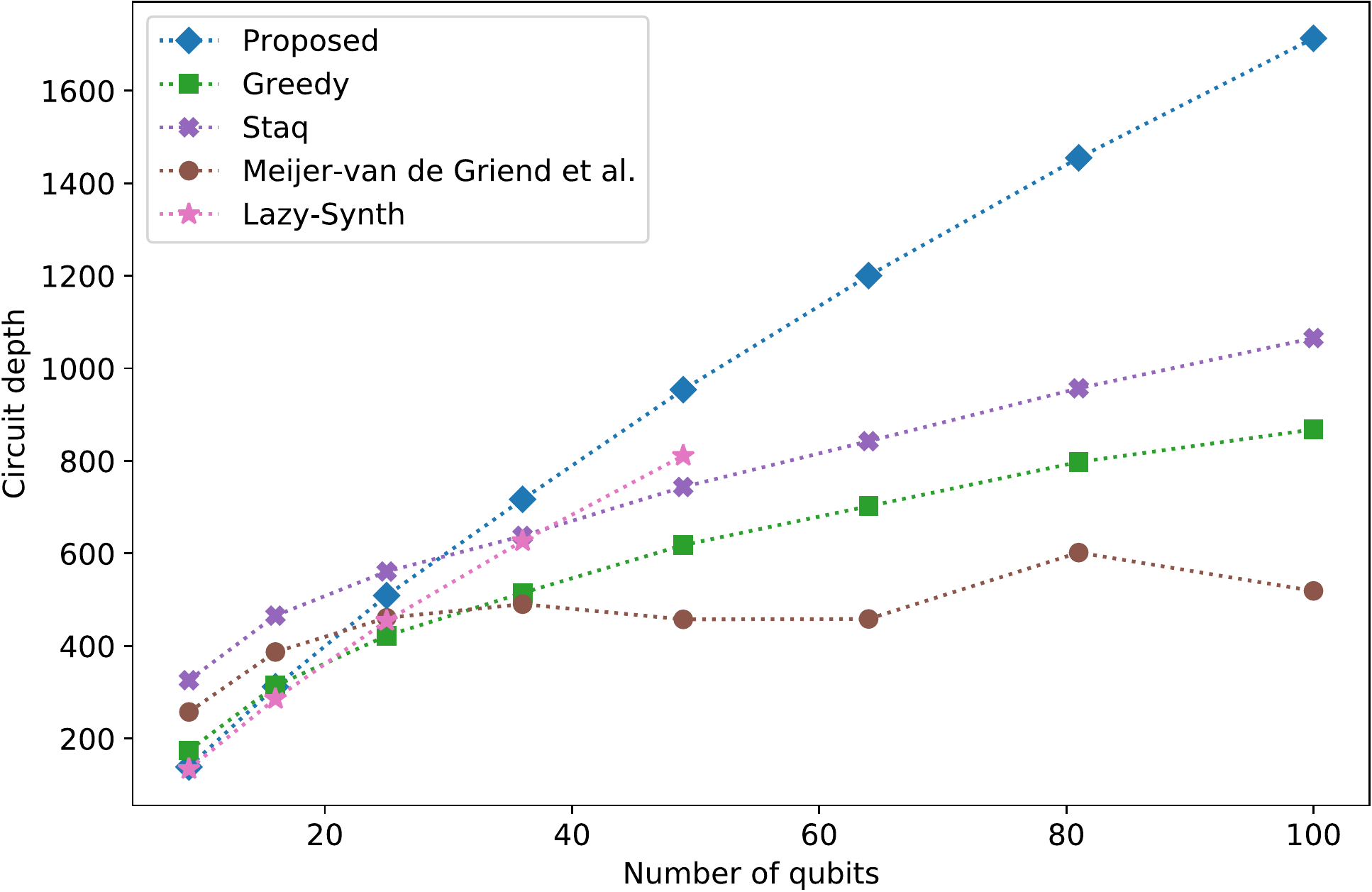}
    \caption{Scaling of the circuit size and circuit depth on square grids for $100$ parities.
    Each point is averaged over $100$ randomly generated parity tables.}
	\label{fig:grid_qubits}
\end{figure}

We also did benchmarks for a constant number of parities with an increasing number of qubits, the results are shown in Figure~\ref{fig:grid_qubits}.
We didn't run the sliding window variants for this benchmark because their usefulness is limited by the small constant number of parities.
We can notice that the algorithm proposed by Meijer-van de Griend et al. has a smaller CNOT depth than the other algorithms when the number of qubits is greater than $36$, however it also has the worst CNOT count.
Algorithm~\ref{code:architecture} and the Lazy-Synth algorithm are producing again the smallest circuits.
Nevertheless, as it can be seen in Table~\ref{tab:alltoall_time_qubits}, they both have a rapidly growing computational time.
The greedy method seems to offer the best compromise.
Indeed, it produces circuits with significantly better CNOT count than the algorithms proposed by Meijer-van de Griend et al. and Staq, and has a much more tolerable running time than Algorithm~\ref{code:architecture} and the Lazy-Synth algorithm.

\begin{table}[H]
    \caption{Average computational time in seconds on square grids for $100$ parities.}
	\label{tab:grid_qubits_time}
	\centerline{
    \scalebox{1.05}{
    \begin{tabular}{|c|cccccccc|}
		\hline
		\multirow{2}{*}{Algorithm} & \multicolumn{8}{c|}{Grid size}\\
		& $3\times 3$ & $4\times 4$ & $5\times 5$ & $6\times 6$ & $7\times 7$ & $8\times 8$ & $9\times 9$ & $10 \times 10$\\
		\hline
			Proposed & 0.303 & 1.439 & 5.763 & 18.650 & 50.745 & 137.250 & 339.595 & 768.073 \\
			Greedy & 0.088 & 0.209 & 0.498 & 1.070 & 2.095 & 4.270 & 8.254 & 15.098 \\
			Staq & \bf 0.038 & \bf 0.078 & \bf 0.157 & \bf 0.268 & \bf 0.421 & \bf 0.648 & \bf 0.990 & \bf 1.616 \\
			Meijer-van de & \multirow{2}{*}{0.066} & \multirow{2}{*}{0.217} & \multirow{2}{*}{0.517} & \multirow{2}{*}{0.941} & \multirow{2}{*}{1.577} & \multirow{2}{*}{2.584} & \multirow{2}{*}{3.960} & \multirow{2}{*}{5.977} \\
			Griend et al. & & & & & & & & \\
			Lazy-Synth & 0.560 & 5.524 & 80.920 & 523.639 & 2381.802 & - & - & - \\
		\hline
	\end{tabular}}}
\end{table}

\section{Conclusion}
We presented heuristic algorithms for the synthesis of circuits over the $\{CNOT, R_Z\}$ gate set.
We covered both cases of full connectivity and partial connectivity for NISQ architectures.
When compared to the state of the art, the benchmarks have shown that our algorithms are producing circuits of smaller or comparable size.
State of the art algorithms yielding analogous CNOT count and CNOT depth performances are outperformed when it comes to the execution time.

Our framework could be further expanded by adding depth and width to our algorithms, which would result in an even smaller CNOT count at the cost of a runtime increase.
This methods was utilized in the Lazy-Synth framework, instead of going in the same direction we presented methods to lower the complexity of our algorithms while preserving as much performances as possible, with the aim of providing a more comprehensive toolkit for the synthesis of phase polynomials.

Finally, the modularity of our approach makes it easily adaptable for the synthesis of other circuits such as sequences of Pauli rotations, a generalization of phase polynomials.
This is not the case for the Gray-Synth algorithm and all its architecture-aware derivatives.
Indeed, the Gray-Synth algorithm relies on the fact that the row $j$ of the parity table isn't affected when applying the gate $CNOT_{i, j}$.
This invariant isn't true when considering sequences of Pauli rotations.
Therefore the Gray-Synth algorithm needs to be coupled with phase polynomials extraction routines in order to compile a generic circuit.

\section*{Acknowledgments}
This work was supported in part by the French National Research Agency (ANR) under the research project SoftQPRO ANR-17-CE25-0009-02, and by
the DGE of the French Ministry of Industry under the research project PIAGDN/QuantEx P163746-484124.
\bibliographystyle{unsrt}
\bibliography{ref.bib}

\end{document}